\documentstyle[12pt]{article}
\def\lb       {\left( }
\def\rb       {\right) }
\def\lmb      {\left\{ }
\def\rmb      {\right\} }
\def\lbb     {\left[ }
\def\rbb      {\right] }
\def\comma      { \, , }
\def\period     { \, . }
\def\bra#1      { \langle \, #1 \, \vert \, }
\def\ket#1      { \, \vert \, #1 \, \rangle \, }
\def\semiket#1  { \, #1 \, \rangle \, }
\def\del        {  \partial  }
\def\half       {  {1\over 2}  }
\def\Tr      {  \mbox{Tr}  }
\def\abs#1      {  \, \vert #1 \vert \,   }
\def\bfR     { {\bf R}}
\def\bfZ     { {\bf Z}}
\def\bfC     { {\bf C}}
\def\calB       { {\cal B} }

\def\vecii#1#2      {  \left(\begin{array}{c}#1\\#2\end{array}\right)  }
\def\veciii#1#2#3   {  \left(\begin{array}{c}#1\\#2\\#3\end{array}\right)  }
\def\matrixii#1#2#3#4            {  \left(\begin{array}{cc}#1&#2\\#3&#4
                                       \end{array}\right) }
\def\matrixiii#1#2#3#4#5#6#7#8#9 {  \left(\begin{array}{ccc}#1&#2&#3\\
                                     #4&#5&#6\\#7&#8&#9\end{array}\right)  }
\def\eqabegin         {  \begin{eqnarray}  }
\def\eqaend           {  \end{eqnarray}  }
\def\nn               {  \nonumber  }

\def\parmedskip        {  \par\medskip  }

\def\parbigskipn        {  \par\bigskip\noindent  }
\def\parmedskipn        {  \par\medskip\noindent  }
\def\parsmallskipn      {  \par\smallskip\noindent  }

\def\sectionnumbering { \setcounter{equation}{0}
         \renewcommand{\theequation}{\arabic{section}.\arabic{equation}}}
\def\appendixnumbering#1 { \setcounter{equation}{0}
         \renewcommand{\theequation}{#1.\arabic{equation}}}
\def\SLtwo {SL(2,{\bf R})}
\def\cSLtwo{\widetilde{SL}(2,{\bf R})}
\def\sltwo {sl(2,{\bf R})}

\def\ginv {g^{-1}}
\def\delbar {\bar{\partial}}
\def\zbar   {\bar{z}}
\def\wbar   {\bar{w}}
\def\Jbar   {\tilde{J}}

\def\Lbar   {\tilde{L}}
\def\Nbar   {\tilde{N}}
\def\hatt   {\hat{t}}
\def\hatphi {\hat{\varphi}}
\def\hatr   {\hat{r}}

\def\hatE  { \hat{E} }
\def\hatN  { \hat{N} }
\def\thetaL {\theta_L}
\def\thetaR {\theta_R}
\def\D#1     {{}^{#1} \! D}

\def\rp     {r_+}
\def\rmi    {r_-}
\def\deltap {\Delta_+}
\def\deltam {\Delta_-}
\def\alp {\alpha'}

\def\Ichi { {\cal I}_\chi }

\def\ketpm#1  { \vert \ #1 \  \rangle {}_{\pm} }
\def\ketp#1 { \vert \ #1 \bigm)  }

\def\tilg  {\tilde{g}}

\begin{document}
%%%%%%%%%%%%%%%%%%%%%
%    cover
%%%%%%%%%%%%%%%%%%%%%
%
%%%%%%%%%%%%%%%%%%%%%%%%%%%%%%%%%%%
\def\papertitlepage{\baselineskip 3.5ex \thispagestyle{empty}}
\def\Title#1{\baselineskip 1cm \vspace{1.5cm}\begin{center}
 {\Large\bf #1} \end{center}
\vspace{0.5cm}}
\def\Abstract {\vspace{1.0cm}\begin{center} {\large\bf Abstract}
           \end{center} \par\bigskip}
\def\preprinumber#1#2#3#4{\hfill \begin{minipage}{4.2cm}  #1
              \par\noindent #2
              \par\noindent #3
              \par\noindent #4 \end{minipage}}
\renewcommand{\thefootnote}{\fnsymbol{footnote}}
%
%%%%%%%%%%%%%%%%%%%%%%%%%%%%%%%%%%%%%%%%%%%%%%%
%
\papertitlepage
\preprinumber{TIT/HEP-345}{UT-Komaba 96-18}{hep-th/9611041}{November 1996}
\Title{ String Theory on Three Dimensional Black Holes }
\begin{center}
   {\sc Makoto ~Natsuume
   \footnote[2]{makoto@th.phys.titech.ac.jp; JSPS Research Fellow} } \\
 \vskip -1ex
  {\sl  Department of Physics, Tokyo Institute of Technology \\
 \vskip -2ex
   Oh-Okayama, Meguro, Tokyo 152, Japan } \\
 \vskip -1ex
    and \\
 \vskip -1ex
   {\sc Yuji ~Satoh
   \footnote[3]{ysatoh@hep1.c.u-tokyo.ac.jp; JSPS Research Fellow } }\\
 \vskip -1ex
    {\sl Institute of Physics, University of Tokyo, Komaba \\
 \vskip -2ex
   Meguro, Tokyo 153, Japan }
\end{center}
%%%%%%%%%%%%%%%%%%%%%%%
\baselineskip=0.6cm
%%%%%%%%%%%%%%%%%%%%%%%
\Abstract
We investigate the string theory on three dimensional black holes
discovered by Ba\~{n}ados, Teitelboim and Zanelli
in the framework of conformal field theory.
The model is described by  an orbifold of the $ \widetilde{SL}(2, {\bf R}) $
WZW model.
The spectrum is analyzed by solving the level matching condition and we obtain
winding modes. We then study the ghost problem and show explicit
examples of physical states with negative norms. We discuss the tachyon
propagation
and the target space geometry, which are irrelevant to the details of the
spectrum. We find a self-dual T-duality
transformation reversing the black hole mass. We also discuss
difficulties in string theory on curved spacetime
and possibilities to obtain a sensible string theory on
three dimensional black holes.
This work is the first attempt to quantize a string theory in a black hole
background with an infinite number of propagating modes.
%
%%%%%%%%%%%%%%%%%%%%%%%
\begin{flushright}
\end{flushright}

\newpage
\renewcommand{\thefootnote}{\arabic{footnote}}
\setcounter{footnote}{0}
\baselineskip = 0.6 cm

\pagestyle{plain}
\setcounter{page}{1}

%%%%%%%%%%%%%%%%%%%%%%%%%%%%%%%%%%%%%%%
%
%%%%%%%%%%%%%%%%%%%%%%
% introduction
%%%%%%%%%%%%%%%%%%%%%%
%
\section{Introduction}
Black holes provide useful laboratories in quantum gravity.
Through the study of black holes, we expect
to obtain useful insights in order to
solve problems such as singularities, black hole thermodynamics and Hawking
radiation. In string theory, most discussions on black hole physics are
based on low energy effective theories, but for definite arguments we have
to develop analysis beyond the $ \alpha' $ expansion.

Many works have been devoted to the $ \SLtwo /U(1) $ black hole
\cite{Witten,MSW}
for that reason.
However, most works are based on the semi-classical
analysis, e.g., \cite{Witten}-\cite{Giveon}
and we need further investigations in order to
clarify important issues in black hole physics.\footnote
{See however \cite{DVV}-\cite{IKOS} for example.}
The difficulties are rooted
in the fact that the target space is non-compact and curved in time
direction.

Such difficulties are not characteristic of string theories in black hole
backgrounds. In general, as a sensible physical theory, a string theory has
to satisfy various consistency conditions.
Although we have many consistent string theories on curved spaces, i.e., on
group manifolds, they are compact and must be tensored with Minkowski
spacetime. We have few consistent string theories with
curved time.
For instance, the no-ghost theorem requires a flat light-cone direction.
Even though most proofs \cite{noghost} are stated for the $D=26$ bosonic
string, many can be extended easily to the general $c=26$ matter CFT with
$D$ dimensional Minkowski spacetime and a compact CFT.
The only assumption needed for the compact CFT is that it is conformally
invariant with the appropriate central charge so that there is a nilpotent
BRST operator, and that it has a positive inner product. However, all known
proofs require $D$ to be at least two.
There is no general result for $D<2$.

Since string theory is regarded as the fundamental theory including
gravity, it is important to construct a consistent string theory on curved
{\it spacetime}. There have been a few previous attempts besides the
$ \SLtwo /U(1) $ case. For example, there are various attempts
using the $ \SLtwo $ WZW model \cite{BOFW}-\cite{BN},
but it is known to contain
ghosts.\footnote{A resolution to the ghost problem has been proposed though
\cite{Bars}.}
Russo and Tseytlin has discussed a string theory in a curved
background which can be transformed to a flat theory by T-duality
\cite{RT}.

The purpose of this paper is to formulate the string theory on the three
dimensional black hole discovered by Ba\~{n}ados, Teitelboim and Zanelli
(BTZ) \cite{BTZ}.
This black hole is important in string theory. This is one of
few known exact solutions in string theory and one of the simplest
solutions; the solution is described by an orbifold of the $ \cSLtwo $ WZW
model \cite{HW,Kaloper}. Moreover, strings in three dimensions have an infinite
number of
propagating modes, so it resembles higher dimensional ones.

The BTZ black hole provides a background to the bosonic string, but
it was originally found as a solution to Einstein
gravity.
In fact, it has been extensively studied in Einstein gravity
(for a review, see \cite{Carlip}).
The BTZ black hole shares many properties
with the $ (3+1) $-dimensional black hole.
But it is simpler since it is locally
three dimensional anti-de Sitter $ (AdS_3) $ space.
This simplicity enables us to investigate many characteristics of
the black hole physics in an explicit manner without mathematical
complications.
In the classical theory, for example,
the gravitational collapse and the instability of the inner horizon have
been studied in detail. Quantum field theory on the BTZ black hole
has been also explored and
exact results are known about Green functions,
mode functions and thermodynamic quantities of
scalar fields. Furthermore,
its thermodynamic and statistical mechanical properties
have been investigated by the
Chern-Simons formulation of the $ (2+1) $-dimensional general relativity.

Therefore, the subject is important
both as a quantum black hole and as a string theory in a nontrivial
background.
Nevertheless,
the detailed construction of the orbifold
has not been made so far. In this
paper, we will investigate the spectrum of the theory, the ghost problem,
the tachyon propagation and the target space geometry. Although we cannot
overcome all the problems, this work may provide useful insights into these
issues. Besides this work is the first attempt to quantize a string theory
in a black hole background with an infinite number of propagating modes.

The organization of the present paper is as follows. In Sec. 2,
we briefly
review the BTZ black hole using the $ \cSLtwo $ WZW model. In Sec. 3, we
develop the conformal field theory for the three dimensional black hole. We
investigate the spectrum by solving the level matching condition. Then
we investigate the issue of ghosts in Sec. 4.  We find
explicit examples of physical states with negative norms.
In Sec. 5, we
study the tachyon propagation and the target space geometry.
We discuss states localized near
the black hole and discuss a T-duality transformation reversing the
black hole mass. In Sec. 6, we discuss the
other consistency conditions in string theory and discuss difficulties
in the case of curved spacetime. Basic properties of the representation
theory of $ \cSLtwo $, which are necessary in the text, are summarized in
Appendix A. Representations in the hyperbolic basis are explained in some
detail. Also, we show the Clebsch-Gordan decomposition of the $ \sltwo $
Kac-Moody module in the hyperbolic basis in Appendix B.
%
%%%%%%%%%%%%%%%%%%%%%
%  section 2
%%%%%%%%%%%%%%%%%%%%%
%
\section{The $\cSLtwo /\bfZ $ black hole }
\sectionnumbering
In this section, we briefly review how to describe the BTZ black hole
\cite{BTZ}
from the $ \SLtwo $ WZW model \cite{HW,Kaloper}
and summarize basic facts on the $ \SLtwo $ WZW model.

\subsection{The BTZ black hole as a string background}
We start with the $ \SLtwo $ WZW model
\footnote{There are difficulties to construct a CFT based on a
non-compact
group manifold. In this paper, we will simply assume the existence of the
$ \cSLtwo $
WZW model.} with action
\eqabegin
&& \frac{k}{8\pi} \int_{\Sigma} d^2 \sigma \sqrt{h} h^{\alpha \beta}
      \Tr \left( \ginv \del_\alpha g \ginv \del_\beta g  \right)
    + i k \Gamma (g) \comma
\eqaend
where $ h_{\alpha \beta } $ is the metric on a Riemann surface $ \Sigma $
and $ g $ is an element of $ \SLtwo $.
$ \Gamma $ is the Wess-Zumino term given by
\eqabegin
   && \frac{1}{12\pi} \int_{B}
      \Tr \left( \ginv d g \wedge \ginv d g \wedge\ginv d g \right)
   \comma
\eqaend
where $ B $ is a three manifold with boundary $ \Sigma $.
We parametrize $ g $ by
\eqabegin
  &&g = \matrixii{x_1 + x_2}{x_3 + x_0}{x_3 - x_0}{x_1 - x_2}
     \comma \\
  && \nn \\
  && {\rm det} \ g  = x_0^2 + x_1^2 - x_2^2 - x_3^2 = 1
      \period
\eqaend
The latter equation is nothing but the embedding equation
of the three dimensional anti-de Sitter space ($ AdS_3 $) in a flat space; thus
$ \SLtwo $ and $ AdS_3 $ are the same manifold. This is
the reason why the BTZ black hole is described by the
$ \SLtwo $ WZW model.

In order to unwrap the compact time direction of $ \SLtwo $,
we go to the universal covering group $ \cSLtwo $
and consider three regions parametrized by
\eqabegin
 \begin{array}{lllll}
\mbox{ Region I} &  ( \hatr ^2 > 1) &:&
   x_1 = \hatr \cosh \hatphi \comma &
     x_0 = \sqrt{\hatr^2 -1} \sinh \hatt \comma
     \\
  && & x_2 = \hatr \sinh \hatphi \comma &
      x_3 = \sqrt{\hatr^2 -1} \cosh \hatt \comma
   \\
\mbox{ Region II} &  ( 1 > \hatr ^2 > 0) &:&
   x_1 = \hatr \cosh \hatphi \comma &
      x_0 = \sqrt{1 -\hatr^2} \cosh \hatt \comma
    \\
 & && x_2 = \hatr \sinh \hatphi \comma &
      x_3 = \sqrt{1 -\hatr^2} \sinh \hatt \comma
   \\
\mbox{ Region III} & ( 0 > \hatr ^ 2 ) &:&
   x_1 = \sqrt{-\hatr^2} \sinh \hatphi \comma &
         x_0 = \sqrt{1 -\hatr^2} \cosh \hatt \comma
    \\
 &&& x_2 = \sqrt{-\hatr^2} \cosh \hatphi \comma &
         x_3 = \sqrt{1 -\hatr^2} \sinh \hatt \comma
 \end{array}
\eqaend
where $ - \infty < \hatt \comma \hatphi < \infty $.
\footnote{Throughout this paper, we use dimensionless coordinates.
The dimension is recovered by the cosmological constant $ -l^{-2} $
if necessary.}
These regions
describe I) the region outside the outer horizon, II) the region between the
outer and the inner horizon, and III) the region inside the inner horizon
of the black hole.
In every parametrization, the WZW action takes the form
\eqabegin
  S &=& \frac{1}{4\pi \alp} \int d^2 \sigma \sqrt{h}
     \left( h^{\alpha \beta} G_{\mu \nu}
       + i \epsilon^{\alpha \beta} B_{\mu \nu} \right)
          \del_\alpha X^\mu \del_\beta X^\nu \comma
\eqaend
where
\eqabegin
 ds^2   &=& \alp k \lmb
     - (\hatr^2 -1) d \hatt^2 + \hatr^2 d \hatphi^2
                       + (\hatr^2 -1)^{-1} d \hatr^2  \rmb \comma
    \nn \\
  B &=&  \alp k  \hatr^2 d \hatphi \wedge d \hatt
   \period \label{GBhat}
\eqaend
We make a further change of variables:
\eqabegin
    \hatr^2 & = & \frac{r^2 - \rmi^2}{\rp^2 - \rmi^2} \comma
                      \quad
  \vecii{\hatt}{\hatphi} =
     \matrixii{\rp}{-\rmi}{-\rmi}{\rp}
   \vecii{t}{\varphi} \label{tphi} \comma
\eqaend
where $ r_\pm $ $ ( \rp > \rmi ) $ are positive constants.
Then, we get
\eqabegin
  d s^2  &=& \alp k \lmb
 - \left( r^2 - M_{BH} \right) d t^2 - J_{BH} d t d \varphi
   + r^2  d \varphi^2
     + \left(  r^2 - M_{BH} + \frac{J_{BH}^2}{4r^2}\right)^{-1} d r^2 \rmb
     \comma \nn \\
   B  &=&
      \alp k \ r^2 d \varphi \wedge d t
   \comma \label{BTZbf}
\eqaend
where $ M_{BH} = \rp ^2 + \rmi ^2 $ and $ J_{BH} = 2 \rp \rmi $.
$ B  $ is defined up to an exact form.
By identifying $ \varphi $ with $ \varphi + 2 \pi $ and
dropping the region $ r^2 < 0 $, we obtain the BTZ black hole.
The coordinates in (\ref{BTZbf}) now take
$- \infty < t < + \infty $,
$0 \leq \varphi < 2 \pi $ and $0 \leq r < + \infty$.
$ \rp $ and $ \rmi $ represent the location of the outer and the inner horizon.
$ M_{BH} $ and $ J_{BH} $ are the mass and the angular momentum of the
black hole respectively.

The non-rotating black hole is obtained by
$ \rmi = 0 $. The extremal black hole is obtained by
$ \rp = \rmi $ in (\ref{BTZbf}) although various
intermediate expressions become singular.
One can show that the above
geometry is a solution to
low energy field equations. Moreover, the exact metric and
anti-symmetric tensor are given by the replacement $ k $ with $ k - 2 $
\cite{BST}, where $ -2 $ is the second Casimir of the
adjoint representation of $ \sltwo $. The cosmological constant is given by
$ -\l^{-2} = -\alpha^{'-1} (k-2)^{-1} $.

\subsection{Chiral currents and the stress tensor}
The $ \cSLtwo $ WZW model has a chiral $\cSLtwo_L \times \cSLtwo_R$
symmetry.
The corresponding currents are given by
\eqabegin
   J(z) &=& \frac{i k}{2} \del g  g^{-1} \comma \quad
   \Jbar (\zbar) \ = \ \frac{i k}{2} g^{-1} \delbar g
  \comma \label{currents}
\eqaend
where $ z = \ e^{\tau + i \sigma } $ and $ \zbar = \ e^{\tau - i \sigma} $.
The currents act on $ g $ as
\eqabegin
  J^a(z) g(w,\wbar) &\sim& \frac{- \tau^a g }{ z - w } \comma \quad
  \Jbar^a(\zbar) g(w,\wbar) \sim \frac{- g \ \tau^a  }{ \zbar - \wbar }
  \label{Jg} \period
\eqaend
Here, we have defined $ J^a $ $ (a = 0,1,2) $ by
$ J(z) = \eta_{ab} \tau^a J^b(z)$ and similarly for $ \Jbar ^a$,
where $ \eta_{ab} = $ diag $ (-1, 1, 1)$.
$ \tau^a $ form a basis of $ \sltwo $ with
the properties
\eqabegin
  && \left[ \tau^a \comma \tau^b \right] = i \epsilon^{a b}_{\ \ c} \ \tau^c
   \comma    \quad
 \Tr \left( \tau^a  \tau^b \right) = - \half \ \eta^{ab}.
\eqaend
In terms of the Pauli matrices,
$\tau^0 = - \sigma^2/2 , \tau^1 = i \sigma^1/2$ and $\tau^2 = i \sigma^3/2$.
The stress tensor is given by
\eqabegin
   T(z) &=& \frac{1}{k-2} \eta_{ab}J^a(z)J^b(z)
  \period
\eqaend
The conformal modes of the currents and the stress tensor satisfy
the commutation relations
\eqabegin
   \left[ J^a_n \comma J^b_m \right]
    &=&  i  \epsilon^{ab}_{\ \ c} J^{c}_{n+m} + \frac{k}{2} n
       \eta^{ab} \delta_{m + n} \comma \nn \\
  \left[ L_n \comma J_m^a \right] &=& - m J^a_{n+m} \comma \\
  \left[ L_n \comma L_m \right] &=& (n-m) L_{n+m} + \frac{c}{12}n(n^2-1)
   \delta_{n+m} \comma \nn
\eqaend
where $ c = 3k/(k-2) $. For the critical value $ c = 26 $, we have
$ k = 52/23 $.
The above Kac-Moody algebra is expressed in the basis
$ I^\pm_n \equiv  J^1_n \pm i J^2_n $ and $ I^0_n \equiv J^0_n $ as
\eqabegin
  \left[ I^+_n , I^-_m \right] &=& -2 I^0_{n+m} +  k n \delta_{n+m}
  \comma \quad \left[ I^\pm_n , I^\pm_m \right] \ = \ 0 \comma \nn \\
  \left[ I^0_n , I^\pm_m \right] &=& \pm I^\pm_{n+m} \comma
  \qquad \qquad \qquad
   \left[ I^0_n , I^0_m \right] \ = \ - \frac{k}{2} n \delta_{n+m}
   \period
\eqaend
On the other hand,
in the basis $ J^\pm_n \equiv J^0_n \pm J^1_n  $ and
 $ J^2_n $, the algebra is written as
\eqabegin
  \left[ J^+_n , J^-_m \right] &=& -2 i J^2_{n+m} - k n \delta_{n+m}
  \comma \quad \left[ J^\pm_n , J^\pm_m \right] \ = \ 0 \comma \nn \\
  \left[ J^2_n , J^\pm_m \right] &=& \pm i  J^\pm_{n+m} \comma
  \qquad \qquad \qquad
   \left[ J^2_n , J^2_m \right] \ = \  \frac{k}{2} n \delta_{n+m}
   \period \label{JJcom}
\eqaend
In this paper, we will utilize the latter basis.
Note the Hermite conjugates for the latter basis are given by
\eqabegin
  && \left( J^\pm_m \right)^{\dag}  = J^\pm_{-m} \comma \quad
     \left( J^2_m \right)^{\dag} = J^2_{-m} \period \label{dagJ}
\eqaend

Similar expressions hold for the anti-holomorphic part.

\subsection{Twisting}
As explained in the previous subsection,
in order to get the
three dimensional black hole,
we have (i) to go to the universal covering space of
$ \SLtwo $,
(ii) to make the identification $ \varphi \sim \varphi + 2 \pi $ and
(iii) to drop the region $ r^2 < 0 $.
We can take (i) into account by considering the representation theory
of $ \cSLtwo $ instead of $ \SLtwo $.
%We will mention this point again in appropriate places in the sequel.
The point (iii) is related to
the problem of closed timelike curves \cite{BTZ,HW};
we will discuss this point in Sec. 6.
For now, we will concentrate on (ii).

{}From the $ AdS_3 $ point of view, the translations of
$ \hatt $ and $ \hatphi $ correspond to boosts in the flat spacetime
in which $ AdS_3 $ is embedded. From (\ref{tphi}),
the translation of $ \varphi $ is given by a linear combination of
those of $ \hatt $ and $ \hatphi $. In terms of the
$ \SLtwo $ WZW model, the translations of $ \hatt $ and $ \hatphi $
correspond to a vector and an axial symmetry of the WZW model \cite{HW}.
If we gauge these symmetries, the resulting coset theories are
the $ \SLtwo / U(1) $ black holes \cite{Witten}.

In order to express these translations by the $ \sltwo $
currents, it is convenient to parametrize the group manifold by
analogues of Euler angles; we parametrize Region I-III by
\eqabegin
\begin{array}{lllllll}
   \mbox{ Region I}
     & \! : \!&
   g
   &\! = \!&  e^{- i \thetaL \tau^2} e^{- i \rho \tau^1}
           e^{-i \thetaR \tau^2}
    &\! = \! &
  \matrixii{ e^{\hatphi} \cosh \rho/2}{ e^{\hatt} \sinh \rho/2}
         { e^{-\hatt} \sinh \rho/2}{ e^{-\hatphi} \cosh \rho/2}
     \comma  \\
  \mbox{ Region II}
   &\! : \! &
  g &\! =\! &  e^{-i \thetaL \tau^2} e^{-i \rho \tau^0}
       e^{-i \thetaR \tau^2}
    & \! = \!&
  \matrixii{ e^{\hatphi} \cos \rho/2}{ e^{\hatt} \sin \rho/2}
         { - e^{-\hatt} \sin \rho/2}{ e^{-\hatphi} \cos \rho/2}
    \comma   \label{gmat} \\
   \mbox{ Region III}
   & \! : \! & g
    & \! = \!
   &  e^{-i \thetaL \tau^2} \ s \ e^{-i\rho \tau^1} e^{-i \thetaR \tau^2}
    & \! = \! &
\matrixii{ e^{\hatphi} \sinh \rho/2}{ e^{\hatt} \cosh \rho/2}
  { - e^{-\hatt} \cosh \rho/2}{ - e^{-\hatphi} \sinh \rho/2}
     \comma
    \end{array}
\eqaend
where $ s = \matrixii{0}{1}{-1}{0} $,
\eqabegin
  \thetaL &=& \hatphi + \hatt \comma \quad \thetaR \ = \ \hatphi - \hatt
   \comma \label{thetaLR}
\eqaend
and
\eqabegin
  \begin{array}{lllll}
   \mbox{ Region I} &:&
     \hatr = \cosh \rho/2 \comma &
   \sqrt{\hatr ^2 - 1} = \sinh \rho/2 \comma
    & ( \rho > 0 ) \comma \\
 \mbox{ Region II} &:&
   \hatr  = \cos \rho/2 \comma &
  \sqrt{1 - \hatr ^2 } = \sin \rho/2 \comma
    & ( \pi > \rho > 0 ) \comma \label{rhor} \\
    \mbox{ Region III} &:&
    \sqrt{- \hatr ^2 } = \sinh \rho/2 \comma &
   \sqrt{1 - \hatr ^2 } = \cosh \rho/2 \comma
    & ( \rho > 0 ) \period
  \end{array}
\eqaend
The currents (\ref{currents}) then take the form, e.g.,
\eqabegin
    J^2 &=&
     \frac{k}{2} \left( \del \thetaL + (2\hatr^2 -1) \del \thetaR \right)
   \comma \qquad
  \Jbar^2  \ = \
    \frac{k}{2}
   \left( \delbar \theta_R + (2\hatr^2 -1) \delbar \theta_L \right)
  \period \label{J2}
\eqaend

The translations of $  \hatt $ and $ \hatphi $
are generated by the linear combinations $ J^2_0 \pm \Jbar^2_0 $
from (\ref{Jg}).
The
translation of $ \varphi $ is generated by
$ Q_{\varphi} \equiv \deltam J_0^2 + \deltap \Jbar^2_0 $,
where $ \Delta_\pm = \rp \pm \rmi $.
In terms of $ \thetaL $ and $ \thetaR $,
$ \delta \varphi = 2 \pi $ with fixed $ t $ is expressed by
\eqabegin
   && \deltap \delta \thetaL = \deltam \delta \thetaR = 2 \pi \deltap \deltam
   \label{deltheta} \period
\eqaend
To describe the black hole,
we have to twist (orbifold) the WZW model
with respect to this discrete action. In the following,
we will call our black hole the $ \cSLtwo / \bfZ $ black hole.
%
%%%%%%%%%%%%%%%%%%%%%%%
%    section 3
%%%%%%%%%%%%%%%%%%%%%%%
%
\section{The spectrum of the $ \cSLtwo /\bfZ $ orbifold }
\sectionnumbering
As a consequence of the identification
$ \varphi \sim \varphi + 2 \pi $,
twisted (winding) sectors arise in this theory.
In this section, we will discuss the spectrum including
the twisted sectors.
In orbifolding, the level matching
is required from
the consistency of string theory, for example, modular invariance and
the invariance under
the shift of the world-sheet spatial coordinate.
In addition, we have to check the other consistency conditions such as
unitarity.
These consistency conditions are closely related
to each other.

One difficulty to construct the orbifold is that the field
$ \varphi $ is not a free field.
We are working in a group manifold, so we cannot use
the argument for flat theories.
However, a similar orbifolding has been discussed
in \cite{GPS} to construct a $ SU(2)/\bfZ _N $ orbifold. We will follow
their argument and solve the level matching condition explicitly.
Since we are dealing with a non-compact
group manifold, there are subtleties
as a sensible string theory. We will return to these issues later.

\subsection{Kac-Moody Primaries in the $ \cSLtwo $ WZW model}
Before discussing the orbifold, let us consider Kac-Moody
primaries in the $ \cSLtwo $ WZW model. Operators are
Kac-Moody primary if they
form irreducible representations of global
$ \cSLtwo _L $ $ \times \cSLtwo _R $ and if they are annihilated by
the Kac-Moody generators $ J^a_n $ and $ \Jbar^a_n $ for $ n > 0 $.
For WZW models, they are also Virasoro primary.
For a compact group, local fields
(wave functions) on the group correspond to Kac-Moody primaries
\cite{GPS,GW}.
Thus, we make an ansatz \cite{DVV,CL}
that the Kac-Moody primary fields are
given by local expressions in the fields $ \theta_L , \theta_R $
and $ \rho $, but do not contain derivatives of these fields.
Hence, they take the form
\eqabegin
  && V \lb \theta_L (z, \zbar) , \theta_R (z, \zbar) ,
       \rho (z , \zbar ) \rb
  \period
\eqaend
Furthermore, we assume \cite{DVV} that the Kac-Moody primary fields lead to
normalizable operators, and that
the CFT inherits the natural inner product of the $ \cSLtwo $ representations.
A complete basis
for the square integrable functions on $ \cSLtwo $ is
known in the mathematical literature. It is given by
the matrix elements of the following unitary representations;
the principal continuous series, the highest
and lowest weight discrete
series \cite{VK,DVV}.
Thus, the objects satisfying our requirements
are the matrix elements of the above unitary
representations and they provide the primary fields in the
$ \cSLtwo $ CFT.
We have summarized useful properties of $ \cSLtwo $ representations
in Appendix A.

Note that our choice of the primary fields corresponds to taking
a unitary $\cSLtwo$ representation as a base of the Kac-Moody module.
Most of our discussion below does not change
even if we start with the other representations at the base including
non-unitary ones as in \cite{DN}.

In representations of $ \cSLtwo $, we have three types of basis.
Let us denote the generators of $ \sltwo $ by $ J^0 $, $ J^1 $ and $ J^2 $.
The bases diagonalizing $ J^0 $, $ J^2 $ and $ J^0 - J^1 $ are
called elliptic, hyperbolic and parabolic respectively.
Since we are interested in the orbifolding related to the action
of $ J_0^2 $ and $ \Jbar_0^2 $, we consider representations
in the hyperbolic basis. This basis has been used in the
study of the Minkowskian $ \SLtwo /U(1) $ black hole \cite{DVV,DN}.
We denote three types of primary fields, i.e., the matrix elements
by
\eqabegin
  \begin{array}{ll}
  \D{P} _{J \pm ,J' \pm}^{\chi} \lb g \rb &
     \mbox{ for the principal continuous series,} \\
  \D{H(L)} _{J,J'}^{j} \lb g \rb &
     \mbox{ for the highest $(H)$ and the lowest $ (L)$ series,}
   \label{matrix}
 \end{array}
\eqaend
where $ j $ labels the value of the Casimir; $ J $ and $ J' $ refer to
the eigenvalue of $ J^2 $. For the principal continuous series,
we have additional parameters, $ 0 \leq  m_0  < 1 $
specifying the representation,
and $ \pm$ specifying the base state. $ \chi $ is the pair $ (j,m_0) $.
Under this construction, the primary fields have the common $j$-value in the
left and right sector.
Note that the spectrum of $ J^2 $
ranges all over the real number, namely
$ J , J' \in \bfR $. For the details, see Appendix A.

\subsection{Primary fields in the $ \cSLtwo / \bfZ $ black hole CFT}
We now turn to the $ \cSLtwo / \bfZ $ CFT.
The currents $ J^2(z) $ and $ \Jbar ^2 (\zbar) $ are chiral and have the
operator
product expansions (OPE)
\eqabegin
 && J^2(z) J^2(0) \sim \frac{k/2}{z^2} \comma \quad
  \Jbar^2(\zbar) \Jbar^2(0) \sim \frac{k/2}{\zbar ^2} \period
\eqaend
So, we represent them
by free fields $ \theta_L^F(z) $ and $ \theta_R^F(\zbar) $ as
\eqabegin
  && J^2(z) = \frac{ k}{2} \del \theta_L^F \comma \qquad
     \Jbar^2(\zbar) = \frac{k}{2} \delbar \theta_R^F \period
\eqaend
The normalization of the fields is fixed by
\eqabegin
   \theta_L^F (z) \theta_L^F (0) \sim + \frac{2}{k} \ln z \comma &&
   \theta_R^F (\zbar) \theta_R^F (0) \sim + \frac{2}{k} \ln \zbar
  \period
\eqaend
The signs are opposite to the usual case due to the negative metric of
the $ J^2 $ direction.
The explicit forms of $ \theta_L^F $ and $ \theta_R^F $ are obtained by
integration of (\ref{J2}).
The local integrability is assured by the current conservation.
We also introduce
$ \theta_L^{NF} (z,\zbar) $ and $ \theta_R^{NF} (z,\zbar) $ by
\eqabegin
  \theta_L (z, \zbar) = \theta_L^F (z) + \theta_L^{NF}(z,\zbar) \comma \qquad
  \theta_R (z, \zbar) = \theta_R^F (\zbar) + \theta_R^{NF}(z,\zbar)
 \period
\eqaend
Note $\theta_L^{NF}$ and $\theta_R^{NF}$ are not free fields.

Now, consider the operator
\eqabegin
  W_n(z,\zbar) & \equiv & \exp \lmb - i\frac{k}{2} n
       \lb \deltam \theta_L^F - \deltap \theta_R^F \rb \rmb
  \label{twop} \comma
\eqaend
where $ n \in \bfZ $.
They have the OPE's
\eqabegin
  \theta_L^F(z) W_n(0,\zbar)
 & \sim & - i n \deltam \ln z \cdot W_n(0,\zbar)
  \comma \nn \\
 \theta_R^F(\zbar) W_n(z,0)
 & \sim &  + i n \deltap \ln \zbar \cdot W_n(z,0)
  \period
\eqaend
%[ $ \theta_L^F \theta_L^{NF} \sim 0 $ ??]
Thus, $ \theta_L^F $ and $ \theta_R^F $ shift
by $ 2 \pi \deltam n $ and
$2 \pi \deltap n $, respectively,
under the translation of the world-sheet coordinate
$\sigma \to \sigma + 2 \pi $, i.e.,
$z \to e^{2\pi i } z $ and $ \zbar \to e^{ - 2\pi i } \zbar $. Hence,
$ \delta \varphi = 2 \pi n $ and $\delta t = 0 $
on $ W_n(z,\zbar) $ under
$ \delta \sigma = 2 \pi $. Thus, $ W_n(z,\zbar) $
expresses the twisting with winding number $ n $.

A general untwisted primary field takes the form (\ref{matrix}).
In our parametrization (\ref{gmat}), it is given by
\eqabegin
 V^{j,0}_{J_L,J_R}(z,\zbar)
 & = & D_{J_L,J_R}^j \lb g'(\rho) \rb \ e^{-i J_L \theta_L - i J_R \theta_R }
  \comma \label{V0}
\eqaend
where we have omitted irrelevant indices of the matrix elements.
The explicit form of $ g'(\rho) $ depends on which region we consider.
Combining the untwisted primary field and the twisting operator,
we obtain the general primary field in the $ \cSLtwo /\bfZ $
black hole CFT:
\eqabegin
  V^{j,n}_{J_L,J_R}(z,\zbar) &=& V^{j,0}_{J_L,J_R}(z,\zbar)
      W_n(z,\zbar)  \comma  \label{Vn} \\
 &=& D_{J_L,J_R}^j \lb g'(\rho) \rb
   \ \exp \lmb - i \lb J'_L \theta_L^F + J_L \theta_L^{NF} +
   J'_R \theta_R^F + J_R \theta_R^{NF} \rb \rmb  \comma \nn
\eqaend
where
\eqabegin
  J'_L &=& J_L + \frac{k}{2}\deltam n \comma \quad
  J'_R = J_R - \frac{k}{2}\deltap n \label{JLRp} \period
\eqaend

\subsection{Level matching}
In the previous subsection, we obtained
primary fields. So, a general vertex operator
has the form
\eqabegin
  && J_N \cdot \Jbar_{\Nbar} \cdot  V^{j,n}_{J_L,J_R}(z,\zbar)
  \comma
\eqaend
where $ J_N $ and $ \Jbar_{\Nbar} $ stand for generic products
of the Kac-Moody generators $ J^a_{-n} $ and $ \Jbar^a_{-n} $ respectively.
The untwisted part depends on $ \theta_L^F $ and $ \theta_R^F $ as
$ \exp ( -i \omega_L \theta_L^F -i \omega_R \theta_R^F ) $
and the full operator as
$ \exp ( -i \omega'_L \theta_L^F - i \omega'_R \theta_R^F ) $, where
\eqabegin
   \omega^{(')}_L &=& J^{(')}_L + i ( N_+ - N_- ) \comma \quad
   \omega^{(')}_R \ = \ J^{(')}_R + i ( \Nbar _+ - \Nbar _- ).
\eqaend
$ N_\pm $ and $ \Nbar _\pm $ are the number of $ J^\pm_{-n} $ and
$ \Jbar ^\pm_{-n} $ respectively. Notice that the commutation relation
(\ref{JJcom}) implies that $ J^\pm_{-n} (\Jbar^\pm_{-n}) $
shifts $ \omega_L (\omega_R) $
by $ \pm i $.  This is one feature
of the representations in the hyperbolic basis.\footnote{
This seems to be contradict
     the Hermiticity of $ J^2_0 (\Jbar ^2_0)$.
      However, this is not the case because
     the spectrum of $ J^2_0 (\Jbar ^2_0)$
     is continuous. Representations of $ \SLtwo $
      in the hyperbolic basis has been described in Appendix A.}
If $ \omega^{(')}_{L,R} $ are complex, the vertex operator cannot be
single-valued on the $ \cSLtwo / \bfZ $ manifold. Thus,
we will focus on the vertex operators with $ N_+ = N_- $ and
$ \Nbar _+ = \Nbar _- $, namely, $ \omega^{(')}_{L,R} = J^{(')}_{L,R}$.

The conformal dimension of the vertex operator is obtained
by the GKO
decomposition of the Virasoro algebra.
Decompose the holomorphic part of
the stress tensor as
\eqabegin
   T (z) &=& T^{\sltwo/ so(1,1)} (\rho, \theta_L^{NF}, \theta_R^{NF})
    + T^{so(1,1)} (\theta_L^F) \comma \nn \\
  && T^{so(1,1)} (\theta_L^F) = + \frac{k}{4} \del \theta_L^F
\del \theta_L^F,
  \quad  T^{\sltwo/ so(1,1)} = T  - T^{so(1,1)} \period
\eqaend
Since $ T^{so(1,1)} $ acts only on $ \theta_L^F $, the weight
with respect to $ T^{so(1,1)} $ is given by
$ \Delta^{so(1,1)} (J'_L) \equiv - J_L^{'2}/k   +
$ (the grade of $ J^2_{-n}$'s ).
So, we have
\eqabegin
  L_0 &=&  \Delta^{\sltwo/so(1,1)} (j,J_L)
   + \Delta^{so(1,1)} (J'_L)  \nn \\
 &=&  \frac{-j(j+1)}{k-2} + \frac{J_L^2 - J_L^{'2}}{k} + N
 \label{L0} \comma
\eqaend
where $ -j(j+1) $ is the Casimir, $ N $ is the total grade of $ J^a_{-n} $'s
and $ \Delta^{\sltwo/so(1,1)} $ is
the weight with respect to  $ T^{\sltwo/ so(1,1)} $.
Here, we have used that
$ L_0 $ is given by the Casimir
plus the total grade for the untwisted sector, namely,
\eqabegin
  \Delta^{\sltwo/so(1,1)} (j,J_L) + \Delta^{so(1,1)} (J_L)
   &=& - \frac{j(j+1)}{k-2} + N \period
\eqaend
Similarly, we obtain
\eqabegin
  \Lbar _0 &=& \frac{-j(j+1)}{k-2} + \frac{J_R^2 - J_R^{'2}}{k}
   + \Nbar \label{Lbar0} \period
\eqaend

We are now ready to solve the level matching condition.
The condition is
\eqabegin
   L_0 - \Lbar _0 &=&
  -   n \left[ \left( \deltam J_L + \deltap J_R \right)
    - \frac{k}{2} n J_{BH}   \right]  + N - \Nbar
   \ \in \ {\bf Z} \label{LM} \period
\eqaend
Furthermore, consider the OPE
of two vertex operators with quantum numbers
$ (n_i,J_{L,i},$$ J_{R,i}) $ $ (i = 1,2) $.
Since $ J_{L,R} $ and
$ n $ are conserved, the level matching condition for the resulting
operator reads
\eqabegin
 &&
- ( n_1 + n_2 ) \ \sum_{i=1}^2 \
  \left[ \left(  \deltam J_{L,i} + \deltap J_{R,i} \right)  -
   \frac{k}{2}  n_i J_{BH}   \right]
 \ \in \ {\bf Z} \period
\eqaend
Therefore, if $ J_{L(R),1(2)} $ and $ n_{1,2} $ satisfy (\ref{LM}),
the closure of the OPE requires
\eqabegin
&& \left(  \deltam J_L + \deltap J_R \right)
   - \frac{k}{2} n J_{BH} \ \equiv \ m
 \ \in \ {\bf Z}  \label{solLM} \period
\eqaend
This is the solution to the level matching condition.

We can check single-valuedness of the vertex operator which satisfies
this condition. Let us denote the
$ \theta_{L,R}^{F,NF} $-dependence of (\ref{Vn})
by $ \exp \left( - i \Theta \right)$ and recall
(\ref{deltheta}).
Then, under $\delta \varphi = 2 \pi $,
\eqabegin
 \delta \Theta
  & = & 2 \pi m + \frac{k}{2} \pi n \left[ \frac{1}{\pi}
    \left( \deltam \delta \theta_L^F - \deltap \delta \theta_R^F  \right)
   + \lb \deltap ^2 - \deltam ^2 \rb  \right]
  \period
\eqaend
Hence, the vertex operator is invariant
under
\eqabegin
 && \delta \theta_L^{NF} = \delta \theta_L^F = \pi \deltam \comma \quad
     \delta \theta_R^{NF} = \delta \theta_R^F = \pi \deltap \period
\eqaend
Single-valuedness is guaranteed in this sense.

In our twisting, only the free field part seems relevant.
In the untwisted sector, only the combinations
$ \theta_{L,R} = \theta_{L,R}^F + \theta_{L,R}^{NF} $ appear, so this
does not matter. On the other hand, for a twisted sector, this is
curious because we were originally considering the orbifolding
with respect to
$ \varphi \sim \varphi + 2 \pi  $ including the non-free part.
However, the non-free part {\it is } relevant in the
above sense. This is related to the Noether current ambiguity
in field theory \cite{GPS}.
In any case, one can take the point of view that we are just considering
possible degrees of freedom represented by the twisting
with respect to $ \theta_{L,R}^F $.

So far we have dealt with a generic value of $ \Delta_\pm $ corresponding
to a rotating black hole. For the non-rotating black hole,
set $ \deltap = \deltam  = \rp $ in the above discussion.
Also, we can formally take the extremal limit $ \deltam \to 0 $ at the
end.
However, the procedure to get the extremal black hole from $ \cSLtwo $ is
different from the non-extremal one \cite{BTZ} and singular
quantities appear in the course of the discussion. Thus we have to
examine whether the extremal limit in our result correctly represents
the extremal limit.

\subsection{Physical states}
Let us turn to discussion on physical states. We use the old covariant
quantization. The base states
corresponding to the primary fields are
written as
\eqabegin
    && \ket{ j; J_L , n } \ket{ j; J_R , n } \period
\eqaend
The Kac-Moody module is built on this base by $ J^a_{-n} $.
Since the eigenvalues of $ J^2_0 $ and $ \Jbar ^2_0 $ should be real in our
case,
the number of $ J^\pm_{-n} $ is restricted as mentioned above.
Then the Kac-Moody module of the holomorphic part
is spanned by states
\eqabegin
  K^a_{-1} K^b_{-1} \cdots K^c_{-1} \ket{j;J_L,n} \comma \label{KMmodule}
\eqaend
where $ K^a_{-1} $  $\ (a = +,-,2) $ are defined by
\eqabegin
  && K^+_{-1} = J^+_{-1} J^-_{0} \comma \qquad K^-_{-1} = J^-_{-1} J^+_{0}
   \comma \qquad K^2_{-1} = J^2_{-1}  \period
\eqaend
The states obtained by acting $ J_0^a $
on the above states are excluded unless they result in the above form.

The physical states consist of the left and the
right part of the form (\ref{KMmodule}) and
satisfy the physical state conditions
\eqabegin
      \lb L_n - \delta_n \rb \ket{ \Psi }
   &=& \lb \Lbar _n - \delta_n \rb \ket{ \Psi }
  \ = \ 0 \quad (n \geq 0) \period
\eqaend
The on-shell condition yields
\eqabegin
  J_L &=& - \frac{k}{4} \deltam n + \frac{1}{\deltam n }
   \left( N - 1   - \frac{j(j+1)}{k-2} \right)  \comma \nn \\
 J_R &=& + \frac{k}{4} \deltap n - \frac{1}{\deltap n }
   \left( \Nbar - 1   - \frac{j(j+1)}{k-2} \right) \comma \label{L0n} \\
  N &=& \Nbar + nm \nn
\eqaend
for twisted sectors $ n \neq 0 $,
and
\eqabegin
  1 &=& - \frac{j(j+1)}{k-2} + N \label{L04} \comma \qquad
  N \ = \  \Nbar \label{L00}
\eqaend
for the untwisted sector. Therefore, for a given $ j $, an arbitrarily
excited state is allowed in twisted sectors.
On the contrary, in the untwisted sector, $j$-value is
completely determined by grade $ N $:
\eqabegin
  j &=& j(N) \ \equiv
  \ \frac{1}{2} \left\{  - 1 - \sqrt{1 + 4(k-2) (N-1)} \right\}
  \comma \label{jN}
\eqaend
where we have chosen the branch Re $j \leq -1/2 $ (see Appendix A).
This result is the same as in the string theory on $ \SLtwo $.
%
%%%%%%%%%%%%%%%%%%%%%%%%%
%    section 4
%%%%%%%%%%%%%%%%%%%%%%%%%
%
\section{Investigation of unitarity}
\sectionnumbering
In the previous section, we discussed the spectrum
of the $ \cSLtwo /\bfZ $ orbifold by solving the level matching condition.
But there are other consistency conditions we must take into account,
and as a result, the spectrum in Sec. 3
may be further restricted.

In this section, we will investigate the ghost problem.
The unitary (ghost)
problem for the sting on $ \SLtwo $ has been discussed
and it is shown to contain ghosts \cite{BOFW},\cite{BN,Bars}.
There are attempts
to get a unitary theory by restricting the spectrum
\cite{Petro}. Also, a unitary $ \SLtwo $ theory
has been proposed using modified currents
\cite{Bars}.

In our case, the analysis of unitarity is
different from the string theory on $ \SLtwo $
due to the existence of winding modes and the use of
representations in the hyperbolic basis. However, we can still utilize
a tool developed for the $ \SLtwo $ theory with a slight modification.
Thus, we will first summarize the argument for the $ \SLtwo $ case.
Then, we will show the non-unitarity of
the string on $ \cSLtwo /\bfZ $ orbifold
by constructing physical states
with negative norms.

\subsection{The unitarity problem of the string on $ \SLtwo $}
Let us briefly review the unitarity problem in the $ \SLtwo $ case
\cite{BOFW},\cite{BN,Bars}.
The holomorphic and the anti-holomorphic
parts are independent in the $ \SLtwo $ WZW model
until we consider the modular properties,
so we focus on the holomorphic part.
In order to study the unitarity problem of the $ \SLtwo $ theory,
it is useful to notice the following facts:
\parbigskipn
1. \qquad The on-shell condition is
   the same as (\ref{L00}) or (\ref{jN}).
\parmedskipn
2. \qquad Let $ V^a $ be an operator satisfying
\eqabegin
  \lbb I_0^a , V^b \rbb &=& i \epsilon^{ab}_{\ \ c} V^c \period
\eqaend
(An example is $ V^a = I^a_{-n}$.) Consider the following states:
\eqabegin
   V^+ I_0^- \ketp{j;m} \comma \qquad  V^- I_0^+ \ketp{j;m} \comma \qquad
   V^0 \ketp{j;m} \comma
\eqaend
where $ \ketp{j;m} $ is an eigenstate with the Casimir
$ \bfC = -j(j+1) $ and
$ I_0^0  = m $ (not necessarily base states).
Assume they do not vanish. Then, by evaluating the matrix elements of
the Casimir operator, we find that these states are decomposed into the
representations of $ \sltwo $ with the $j$-values $ j $ and $ j \pm 1 $.
\parmedskipn
3. \qquad
As a consequence of 2),
acting $ I^a_{-1 }$ $ N $ times on a base state $ \ket{j;m} $
yields $ 3^N $ independent states at grade $ N $ with $j$-values
ranging from $ j-N$ to $j+N$.
Let us call the states with $ j\pm N $
the ``extremal states" and denote them by $ \ket{E^\pm_N} $.
$ \ket{E^\pm_N} $ are physical if they satisfy the on-shell condition.
The reason is simple. Since the Casimir
commutes with $ L_n $, $ L_n \ket{E^\pm_N} $ have the same $j$-values as
$ \ket{E^\pm_N} $. However, $ L_n \ket{E^\pm_N} $ are at grade
$ N-n $, and thus
their $j$-values should range from $ j - (N-n) $ to $ j + (N-n) $.
Therefore,
we have $ L_n \ket{E^\pm_N} = 0 $ $\ (n >0)$;
together with the on-shell condition, they are physical.
\parmedskipn
4. \qquad Let $ \ket{\Psi} $ be a physical state, i.e.,
$ (L_n - \delta_n ) \ket{\Psi} = 0 $ for $ n \geq 0 $.
Then the states obtained by acting
$ J^a_0 $ on $ \ket{\Psi} $ are also physical:
\eqabegin
  L_n J_0^a \cdots J_0^b \ket{\Psi} &=& \lbb L_n ,
J_0^a \cdots J_0^b \rbb \ket{\Psi}
  = 0 \period
\eqaend
\parmedskipn
5. \qquad For the discrete series, we have a simple expression of the extremal
states, e.g.,
\eqabegin
  \ket{E^{d+}_N} & = & \lb I^+_{-1} \rb ^N \ket{j(N);j(N)} \comma
  %\label{discext}  %\comma
\eqaend
where $ \ket{j(N);j(N)} $ is a highest-weight state, namely
$ I^+_0 \ket{j(N);j(N)} = 0 $.
It is easy to obtain the norm of this state:
\eqabegin
    \bra{E^{d+}_N} \semiket{ E^{d+}_N} &=&  \bra{j(N);j(N)} \semiket{j(N);j(N)}
      (N !) \prod_{r=0}^{N-1} ( k + 2 j(N) + r ) \label{normE} \period
\eqaend
\parmedskip

We can immediately find physical states with negative
norms.
First, let us consider the case $ k < 2$.
{}From 1) and 3),  $ \ket{E^+_N} $ with $ j = j(N) $
at its base is a physical state.
At sufficiently large $ N $, $j(N)$
takes a value of the principal continuous series. On the other hand, the
$j$-value
of $ \ket{E^{+}_N} $ is $ j(N) + N $. There is no unitary representations
with this $j$-value. Thus, the module
$ I_0^a \cdots I_0^b \ket{E^+_N} $ is physical, but forms a {\it non-unitary}
representation of $ \sltwo $.

Second, let us consider the case $ k > 2 $. Again
$ \ket{E^{d+}_N} $
with $ j = j(N) $ at its base is a physical state.
We easily find that
\eqabegin
  && I_0^+ \ket{ E^{d+}_N }  = 0 \comma \qquad
   I^0_0 \ket{ E^{d+}_N } = \lb j(N) + N \rb \ket{ E^{d+}_N } .
\eqaend
Thus $ \ket{E^{d+}_N} $ is  a highest-weight state of
a highest-weight $ \sltwo $ representation like $ \ket{j(N) + N;j(N) +N} $.
However, the $ I^0 $-value becomes positive for large $ N $ .
Since there is no unitary representation of $ \sltwo $ with such a
highest weight state,
the states in the $ \sltwo $ representation built on
$ \ket{E^+_N} $ by $ I^a_0 $
are physical but some have negative norms.

Although we can flip the sign of the norm of
$ \ket{ j(N);j(N) } $ so that $ \bra{ E^{d+}_N } \semiket{ E^{d+}_N } > 0 $
for arbitrary $ N $, we cannot remove physical states with negative norms.
This is because we have infinitely many physical states built on
$ \ket{ E^{d+}_N } $ as in 4),
and they form a non-unitary $ \sltwo $ representation.

\subsection{Physical states up to grade 1}
Now let us discuss the $ \cSLtwo /\bfZ $ orbifold case.
One difference from the previous discussion is the existence
of winding modes. Thus, for twisted sectors,
(\ref{jN}) does not hold
and  the holomorphic and anti-holomorphic part are not independent.
The other important difference is that
we use the hyperbolic basis and the Kac-Moody module
is restricted to the form (\ref{KMmodule}). We do not
have states given in 4) and 5) in the previous subsection.
Nevertheless, the argument on extremal
states is still valid,
so we will use them to show that our theory is not unitary.

First, let us consider physical states up to grade one.
For the time being, we focus on the holomorphic part.
At grade one, we have three states for a fixed  $j$-, $ J^2_0 $- and
$n$- value,
namely,
\eqabegin
   &&
  \ket{ \pm } \equiv  K^\pm_{-1} \ket{j;\lambda,n} \comma \qquad
   \ket{ 2 } \equiv K^2_{-1} \ket{j;\lambda,n} \period  \label{+-2}
\eqaend
Using the Hermiticity (\ref{dagJ}) and (\ref{Jpm}),
we get norms among
the above states
\eqabegin
  \veciii{ \bra{+} }{ \bra{-} }{ \bra{2} }
   \lb \ket{+} \comma \ket{-} \comma \ket{2} \rb  &=&
  \matrixiii{0}{-(2i\lambda+ k') A}{- i A}{ (2i\lambda -  k') \bar{A}}
{0}{i \bar{A}}{i \bar{A}}{- i A}{k/2}
  \comma
\eqaend
where $A = j(j+1) + \lambda (\lambda + i)$ and $ k' = k-2 $. We have
omitted $ \bra{ j;\lambda,n } \semiket{ j;\lambda,n } $.
These states are decomposed into the eigenstates
of the Casimir with $j$-values $ j $ and $ j \pm 1 $
from the argument in Appendix B.
We denote them
by $ \ket{ \Phi^j (j;\lambda,n) } $ and
$ \ket{ \Phi^{j\pm1}(j;\lambda,n) } $. Note
$ \ket{ \Phi^{j\pm1} } $ are extremal states.
Explicitly, they are
given by (up to normalization)
\eqabegin
  \veciii{ \ket{ \Phi^{j+1} } }{ \ket{ \Phi^{j} } }{ \ket{ \Phi^{j-1} } }
  & \! \! = \! \! &
\matrixiii{j+1-i \lambda }{-(j+1 + i \lambda)}{2i \lb (j+1)^2 + \lambda^2 \rb}
{1}{1}{-2 \lambda}{-(j+i \lambda)}{j-i\lambda}{2i \lb j^2 + \lambda^2 \rb}
      \veciii{ \ket{+} }{ \ket{-} }{ \ket{2} }
   \period
\eqaend

At grade one, the conditions $ L_n = 0 \ \ ( n > 0 ) $  reduce to
$ L_1 = 0 $.
This imposes one equation on a state given by a linear combination
of $ \ket{ \pm } $ and $ \ket{2} $.
Then the space of the solution has (complex)
two dimensions at a generic value of $ j $ and $ \lambda $.
Since
extremal states are physical and we have two extremal states at grade one,
the physical states take the form
\eqabegin
   \alpha \ket{\Phi^{j+1}} + \beta \ket{\Phi^{j-1}}
   \period \label{phys}
\eqaend
At special values of $ \lambda $ and $ j $, we have extra solutions.
Similarly, we can get the states at grade one satisfying the
$ \Lbar_1 = 0 $ condition. Hence, the physical states up to grade one
are obtained by tensoring the holomorphic and the anti-holomorphic
part using the states (\ref{phys})  and
base states which satisfy the on-shell condition
(\ref{L0n}) or (\ref{L00}).
\subsection{Non-unitarity of the string on $ \cSLtwo /\bfZ $ orbifold}
Finding
physical states with negative norms is easy using the above physical states.
First, let us discuss the case of real $ j $
(the discrete series).\footnote{This argument is also valid for
the complementary series although we have not included
this series.}
There exist the following physical states:
\eqabegin
 && \ket{ \Psi^d_1 } = \ket{ j;J_{L,1},1 }  \ket{j;J_{R,1},1}  \comma \quad
   \ket{ \Psi^{d}_2 } = \ket{ \Phi^{j+1} (j;J_{L,2},1) }  \ket{ j;J_{R,2},1 }
 \comma
\eqaend
where $ m_1 = 0 $, $ m_2 = 1 $
and
\eqabegin
   \begin{array}{ll}
    {\displaystyle J_{L,1} =  - \frac{k}{4} \deltam  - \frac{1}{\deltam }
   \left( 1   + \frac{j(j+1)}{k-2} \right) \comma }  &
    {\displaystyle J_{R,1}  =    \frac{k}{4} \deltap  + \frac{1}{\deltap  }
   \left( 1   + \frac{j(j+1)}{k-2} \right) \comma } \\
  {\displaystyle J_{L,2} =   - \frac{k}{4} \deltam  - \frac{1}{\deltam }
    \frac{j(j+1)}{k-2}  \comma } &
  {\displaystyle J_{R,2} \ = \ J_{R,1} \label{JLR} \period }
 \end{array}
\eqaend
By explicit calculation, norms of these states are
\eqabegin
 \bra{ \Psi^d_1 } \semiket{ \Psi^d_1 } & = &
    \bra{ j;J_{L,1},1 } \semiket{ j;J_{L,1},1 }
    \bra{ j;J_{R,1},1 } \semiket{ j;J_{R,1},1 }  \comma \\
  \bra{ \Psi^{d}_2 } \semiket{ \Psi^{d}_2 } & = &
   2  (j+1)(2j+1)(2j+k) \lb (j+1)^2 + J_{L,2}^2  \rb
   \nn \\
&& \qquad  \times
    \bra{ j;J_{L,2},1 } \semiket{ j;J_{L,2},1 }
    \bra{ j;J_{R,2},1 } \semiket{ j;J_{R,2},1 }
        \period \nn
\eqaend
If the bases of $ \ket{ \Psi^d_1 } $ and $ \ket{ \Psi^{d}_2 } $
belong to the same representation of $ \sltwo $, \\
 $  \bra{ j;J_{L,i},1 } \semiket{ j;J_{L,i},1 }
      \bra{ j;J_{R,i},1 } \semiket{ j;J_{R,i},1 }
$  $ (i = 1,2) $ take the same value.
Then, for a sufficiently large $ \abs{ j } $ (recall $ j \leq -1/2 $),
the latter norm behaves as $ 8 j^7/(k'\deltam)^2 $, and
the two norms have opposite signs. Thus, if we include
the
bases with real $ j $,
our orbifold cannot be unitary.

Next, let us turn to the case of complex $ j $ (the principle
continuous series). Because $ j = -1/2 + i \nu \ (\nu > 0)$, the extremal
states at grade one have $ j = -1/2 \pm 1 + i \nu $.
These correspond to complex Casimirs and non-unitary $ \sltwo $
representations. This is not the end of the story however because
(i) infinite series of states build on these states
by the current zero-modes are not
allowed and (ii) the left and right sector are connected by the quantum
numbers $ n $ and $ m $.
In this case, the norm of $ \ket{ \Psi^{d}_2 } $ vanishes.
(Thus we have
infinitely many physical states with zero norm; see Appendix B.)
Hence, consider the following physical states instead:
\eqabegin
  && \ket{ \Psi^p_1 } = \ket{ j;J^1_L,1 }  \ket{ j;J^1_R,1 }  \comma
   \nn \\
 && \ket{\Psi^{p}_2} =
  \lb \ket{ \Phi^{j-1} (j;J_{L,2},1) }  - i \
   \ket{ \Phi^{j+ 1} (j;J_{L,2},1)}  \rb
   \ket{ j;J_{R,2},1 }
 \comma
\eqaend
where $ J_{L(R),i} $ are given by (\ref{JLR}).
The norms of these states are
\eqabegin
  \bra{ \Psi^p_1 } \semiket{ \Psi^p_1 } &=&
    \bra{ j;J_{L,1},1 } \semiket{ j;J_{L,1},1 }
    \bra{ j;J_{R,1},1 } \semiket{ j;J_{R,1},1 }  \comma \\
  \bra{ \Psi^{p}_2 } \semiket{ \Psi^{p}_2 } &=&
    - 4 \nu \lbb \lb J_{L,2}^2 - 1/4  - \nu^2  \rb
  \lb 4 \nu^2 - 3 k -1  \rb
     + 2 (1+k) J_{L,2}^2 - k  \rbb \nn \\
   &&  \qquad \qquad \qquad \times
    \bra{ j;J_{L,2},1 } \semiket{ j;J_{L,2},1 }
    \bra{ j;J_{R,2},1 } \semiket{ j;J_{R,2},1 }
    \period \nn
\eqaend
For a sufficiently large $ \nu $, the latter norm
behaves as $ -16 \nu^7/(k'\deltam)^2 $. Thus,
the two norms have opposite signs if
the bases of $ \ket{ \Psi^p_1 } $ and $ \ket{ \Psi^{p}_2 } $
belong to the same representation of $ \sltwo $.
Therefore if we include the
bases belonging to the principal continuous series,
our orbifold is again non-unitary.

Notice that bases with large $ \abs{ j } $ or $ \nu $
are generated from those of small values by tensor products
(see Appendix A) unless they decouple.

For the $ \SLtwo $ theory, a physical state at a sufficiently
high grade has large $ \abs{ j } $ at the base, which caused the trouble.
In our case, some ghosts in the $ \SLtwo $ theory disappear,
but physical states with large $ \abs{ j } $ at the base exist
already at grade one due to the winding modes.
We have still possibilities that the orbifold becomes ghost-free,
for instance, by some truncation of the spectrum.
We will discuss this issue in Sec. 6.
%
%%%%%%%%%%%%%%%%%%%%%
%    section 5
%%%%%%%%%%%%%%%%%%%%%
%
\section{Tachyon and target-space geometry}
\sectionnumbering
In this section, we discuss
the tachyon propagation on the $ \cSLtwo/\bfZ $ black hole and
the target-space geometry, which are irrelevant to the details
of the full spectrum.

{}From the group theory point of view, the $ \SLtwo /U(1) $ black hole
and the $ \cSLtwo/\bfZ $ black hole are closely related.
For example, primary fields in both theories
are constructed from the matrix elements of $ \SLtwo $.
Hence we observe similar properties for the tachyon and the
target-space geometry in two theories.

\subsection{Tachyon in the untwisted sector}
First, let us consider the tachyon in the untwisted sector.
It is expressed by the matrix elements of $ \cSLtwo $ in
various unitary representations as (\ref{V0}).
The matrix elements satisfy the differential equation \cite{VK}
\eqabegin
  \lbb  \Delta
   - j (j +1) \rbb  D^{j(\chi)}_{J_L,J_R} \lb g \rb
   &=& 0
  \comma \label{laplace}
\eqaend
where
$ \Delta $ is the Laplace operator on $ \SLtwo $.
Because the geometry of the black hole is
locally $ \SLtwo $, this equation is nothing
but the linearized tachyon equation \cite{GL}
or the Klein-Gordon equation in the BTZ black hole background \cite{IS}
up to a factor.
Then the analysis of the tachyon scattering
and the Hawking radiation in \cite{GL} is valid without change.
We do not
repeat it here, but only make an explicit correspondence between
the untwisted tachyon in \cite{GL} and ours.
$ j(j+1) $ represents the mass-squared. For the untwisted
sector, the on-shell condition is (\ref{L00}) with $ N= \Nbar = 0 $,
so gives
a principal continuous series.

In \cite{GL}, the tachyon is expanded as
\eqabegin
    T &=& \sum_{N \in \bfZ} \int dE \  T_{E N} (r)
    \ e^{-i E t} e^{-i N \varphi } \ = \
   \sum_{\hatE , \hatN} T_{\hatE \hatN } (\hatr)
    \ e^{-i \hatE \hatt} e^{-i\hatN \hatphi } \period
\eqaend
After changes of variables to $ z = 1- \hatr ^2 $ and
\eqabegin
   T_{\hatE \hatN} (z) &=&
  z^{i \hatE/2} ( 1- z)^{i \hatN/2 } \Psi_{\hatE \hatN} (z)
  \comma
\eqaend
we find that $ \Psi_{\hatE \hatN} (z) $ is given by the hypergeometric
function. Comparing the above expression with (\ref{V0}), we get the
correspondences
\eqabegin
   \begin{array}{lcllcl}
   T  & \leftrightarrow  &
       \D{P} ^\chi_{J_L \pm,J_R \pm}(g) \comma \qquad  &
   T_{ \hatE \hatN }  & \leftrightarrow  &
    \D{P} ^\chi_{J_L \pm,J_R \pm}(g')\comma   \\
     J_L + J_R   & \leftrightarrow & \hatN \comma &
     J_L - J_R   & \leftrightarrow &  \hatE \comma
   \end{array}
\eqaend
where we have used (\ref{thetaLR}). Since $ \varphi $ has period
$ 2 \pi $, $ N = - \rmi \hatE + \rp \hatN \in \bfZ $.
This is the level matching condition (\ref{solLM}) with
$ n = 0 $.

As a further check, let us consider the matrix elements for
$ g' =
\matrixii{ \cosh \rho/2 }{ \sinh \rho/2 }{ \sinh \rho/2 }{ \cosh \rho/2 } $
$ \ ( \rho > 0 ) $; this corresponds to the region $ r > \rp $.
They are given by
\eqabegin
  \D{P} ^\chi_{J_L +,J_R +} (g') & = & \frac{1}{2\pi} B(\mu_L,-\mu_L - 2j)
 \frac{\cosh^{2j + \mu_L + \mu_R} \rho/2 }{ \sinh^{\mu_L + \mu_R } \rho/2 }
   \ F \lb \mu_L, \mu_R; - 2 j ; - \sinh^{-2} \rho /2 \rb ,
  \nn  \\
 \D{P} ^\chi_{J_L -,J_R -} (g') & = &
  \frac{1}{2\pi} B\lb 1-\mu_R,\mu_R- 1 + 2(j+1) \rb
\frac{\cosh^{2j + \mu_L + \mu_R} \rho/2 }{ \sinh^{4 j + 2 +  \mu_L + \mu_R }
\rho/2 }
    \label{D--} \\
   && \qquad \quad \times  F \lb \mu_L+ 2j + 1 , \mu_R + 2j + 1 ;
 2 j + 2 ; - \sinh^{-2} \rho /2 \rb \nn \comma
\eqaend
where $ \mu_{L,R} = i J_{L,R} - j $. $ F $ and $ B $ are the
hypergeometric function and the Euler beta function respectively.
Noting $ - \sinh^2 \rho/2 = 1-\hatr^2 \ = \ z $,
we find that $ {}^PD^\chi_{J_L +,J_R +} (g') $ and
$  \D{P} ^\chi_{J_L -,J_R -} (g') $ are
the mode functions
in \cite{GL} which are regular at infinity.
Generically, the untwisted tachyon behaves as
\eqabegin
  \begin{array}{llll}
    \D{P} ^\chi_{J_L +,J_R +} (g') & \sim & a_1 (r^2)^j
    & \mbox{as  } r \to \infty  \comma \\
    & \sim & b_1 \ e^{i(J_L-J_R) \ln \sqrt{r^2 - \rp ^2}}
             + b_2 \ e^{- i(J_L-J_R) \ln \sqrt{r^2 - \rp ^2}}
    & \mbox{as  } r \to \rp
     \comma \\
  \D{P} ^\chi_{J_L -,J_R -} (g') & \sim & a'_1 (r^2)^{-(j+1)}
    & \mbox{as  } r \to \infty
    \comma \\
    & \sim & b'_1 \ e^{i(J_L-J_R) \ln \sqrt{r^2 - \rp ^2}}
             + b'_2 \ e^{-i(J_L-J_R) \ln \sqrt{r^2 - \rp ^2}}
    & \mbox{as  } r \to \rp
   \comma \label{asymptD}
   \end{array}
\eqaend
where $ a^{(')}_{1} $ and $ b^{(')}_{1,2} $ are some constants.
Since Re $ j = -1/2 $,
they behave like spherical waves asymptotically.
When $ J_L = J_R $, hypergeometric functions degenerate;
then the asymptotic behaviors as $ r \to \rp $ are different from
(\ref{asymptD}).

\subsection{Tachyon in twisted sectors}
Now let us turn to the tachyon in twisted sectors.
The twisted tachyon is given by the product of the matrix element and
the twisting operator as (\ref{Vn}).
The twisting operator gives a phase to the tachyon.
In the twisted sectors, various $ j$-values are allowed from the on-shell
condition (\ref{L0n}) with $ m = N =\Nbar = 0 $ and $ n \neq 0$.
Thus the matrix elements of the discrete series appear as well as those of
the principal
continuous series. For the principal continuous series,
the explicit forms and asymptotic behaviors of the matrix elements
are the same as in the untwisted
sector (although $j$-values are different).

For the discrete series,
only one linear combination of solutions to (\ref{laplace}) appears.
As explained in Appendix A, the matrix elements are obtained
from one of the matrix elements in the principal continuous series;
\eqabegin
  \D{L} ^j_{J_L,J_R } (g') & \propto  & \D{H} ^j_{J_L,J_R} (g') \ \propto \
  \D{P} ^\chi_{J_L +,J_R +} (g')  \label{DLH} \period
\eqaend
Thus we can read off the behaviors of $ {}^{L,H}D^j_{J_L,J_R } (g') $
from $ {}^PD^\chi_{J_L +,J_R +} (g') $. Note in particular that
$ {}^{L,H}D^j_{J_L,J_R } (g') \to (r^2)^j$ and $ j \leq -1/2 $. Therefore,
a tachyon state in the discrete series dumps rapidly as one goes to infinity,
so this is a state
localized near the black hole. This is similar
to a winding state in the Euclidean $ \SLtwo /U(1)$ black hole
where one can regard it as a bound state in the dual geometry \cite{DVV}.
Hence, we have two kinds of tachyon.
One is from the principal continuous series and propagates like a wave,
and the other
is from the discrete series and is localized near the black hole.

The differential
equation (\ref{laplace}) for the discrete series
is again the Klein-Gordon equation.
The thermodynamic
properties of the corresponding scalar fields are discussed
in \cite{IS}.

\subsection{Global properties}
So far we have not discussed global properties of the tachyon, but
considered the tachyon propagation in one patch of the orbifold
(the region $ r > \rp $).
In order to define the tachyon propagation globally, we have to continue
it  from one region to another. Let us start with a tachyon
in the region $ r > \rp $. Recall that the regions have the boundaries at
the inner
and the outer horizon ($ r = r_\pm $).
The tachyon is given by a linear combination of
(\ref{D--}) or (\ref{DLH}) and is regular at infinity.
{}From the linear transformation formulas of hypergeometric functions,
we can obtain the expression around $ r = r_\pm $ as in (\ref{asymptD}).
We would like to continue it to the other regions.

Here we have two possible sources of obstacles. One is complex
 power of  $ z $ or $ 1 - z $. This causes troubles
as $ z \to 0 $ ($ r \to \rp $) or $ z \to 1 $ ($ r \to \rmi $).
The other is logarithmic singularities like $ \log z $ and $ \log (1-z) $.
The logarithmic singularity
at $ z = 0 $ ($ r = \rp $) arises when $ \mu_L - \mu_R \in \bfZ $, i.e.,
$ J_L - J_R = 0 $, and the one
at $ z = 1 $ ($ r = \rmi $) arises when $ \mu_L + \mu_R + 2j \in \bfZ $, i.e.,
$ J_L + J_R = 0 $.
The latter corresponds to the case
of the $ \SLtwo/U(1) $ black hole in which the tachyon develops
a logarithmic singularity at the origin (singularity) \cite{DVV}.
This is natural because the inner horizon of the $ \cSLtwo/\bfZ $ black hole
and the origin of the $ \SLtwo/U(1)$ black hole are the same point in
the $ \SLtwo $ group manifold.

Note that the matrix elements are continuous all over the group
manifold. Thus if we consider a generalized function space including
distributions, we can continue the tachyon from one region to
another in any case.
We leave as open problems precise prescription of the continuation
and the physical interpretation of
the above singularities.

\subsection{T-duality}
The
$ \cSLtwo /\bfZ $ black hole has two Killing vectors $ \del_{\hatt} $
and
$ \del_{\hatphi} $. In coordinates $ (\hatt, \hatphi, \hatr) $,
the geometry is given by (\ref{GBhat}) and the dilaton
$ \phi = 0 $.
In order to deal with a general T-duality transformation,
let us define new coordinates $ x $ and $ y $ by
\eqabegin
  \vecii{\hatt}{\hatphi} &=& \matrixii{\alpha }{\beta}{\gamma}{\delta}
              \vecii{x}{y} \comma \quad \alpha \delta - \beta \gamma \neq 0
   \period
\eqaend
Then, the T-duality transformation to $ x $ covers
all T-duality transformations.

First, let us consider the T-duality transformation
to $ \varphi $. This is discussed in \cite{HW}.
Setting $ x = \varphi $
and $ y = t $, the dual of the $ \cSLtwo /\bfZ $
black hole becomes in general the black string.
The T-duality transformation is not self-dual.

Next, let us set $ x = \hatphi $ and $ y = \hatt - \hatphi $. In
these coordinates, the geometry is given by
\eqabegin
    ds^2 &=& \alp k \lmb  dx^2
   + (1 - \hatr^2 ) dy^2
    + 2  ( 1 - \hatr ^2 ) d x d y
          +  ( \hatr ^2 - 1 )^{-1} d \hatr ^2 \rmb \comma \nn \\
    B  &=& \alp k \ \hatr ^2  d x \wedge d y
    \comma \qquad \phi \ = \ 0 \period
\eqaend
The T-duality transformation
\cite{Buscher,GRV}
gives the following dual geometry:
\eqabegin
 \tilde{ds}^2  &=&  \alp k \lmb dx^2
   + \hatr^2  dy^2
    + 2  \hatr ^2  d x d y
          +  ( \hatr ^2 - 1 )^{-1} d \hatr ^2 \rmb \comma \nn \\
   \tilde{B} &=& \alp k ( 1- \hatr ^2 ) d x \wedge d y
    \comma \qquad \tilde{\phi} \ = \ 0
   \period
\eqaend
This geometry is obtained from  the original one via
$ \hatr ^2 \to 1 - \hatr ^2 $, or $ \hatt \leftrightarrow \hatphi $.
Thus, this T-duality transformation
is self-dual and interchanges the outside of the inner horizon\footnote
{or the outside of the origin
for the non-rotating black hole} $ (\hatr ^2 > 0) $ and
the inside of the outer horizon $(\hatr ^2 < 1) $. In particular,
the  outer and the inner horizon (or the origin) are interchanged.
Recall that translations of $ \hatt $ and $ \hatphi $ are the
vector and the axial symmetry. So,
the transformation $ \hatt \leftrightarrow \hatphi $ corresponds
to the T-duality transformation in the $ \SLtwo /U(1) $ black hole
which interchanges the horizon and the singularity
\cite{DVV,Giveon}.

Since $ \varphi $
is periodic, we have to further
specify the periodicity of the dual coordinate.
In the above T-duality transformation, the period of $ x = \hatphi $ in the
dual geometry should be reciprocal of that in the original geometry
\cite{GRV}.
{}From (\ref{tphi}), we see that the periods of $ \hatt $ and $ \hatphi $
are not independent, so generically,
we cannot specify the period of $ \hatphi $ only.
However, for the non-rotating black hole ($ \rmi = 0 $), we have
$ \hatphi = \rp \varphi $ and the period of $ \hatphi $ in the original
geometry is equal to $ 2 \pi \rp $. Hence the period in the dual geometry
is $ 2 \pi/ (\rp k) $. This indicates that the black hole mass
is reversed under the T-duality transformation because $ M_{BH} = \rp ^2 $.
Because $ J_{L,R} $ take all real values, the spectrum of
$ L_0 $ and $ \Lbar _0 $ is
formally invariant under this T-duality transformation. But it is not bounded
from
below as in Minkowski spacetime, so we need some procedure
such as the Wick rotation for a rigorous argument.
%
%%%%%%%%%%%%%%%%%%%%%%%%
%   discussion
%%%%%%%%%%%%%%%%%%%%%%%%
%
\section{Discussion}
\sectionnumbering
\subsection{Consistency conditions}
In Sec. 4, we found that there are physical states with negative norms.
We can speculate various reasons why ghosts survive in our analysis:
\begin{enumerate}
\item  Further truncation might be necessary on the spectrum.
\item Modular invariance might fix the problem.
\item The theory on $ \SLtwo $ might be sick. The $ \SLtwo $ WZW model
describes anti-de~Sitter space, so has unusual asymptotic properties.
\item One might has to use modified currents.
\item We might have to include non-unitary representations for base
representations of current algebras.
\end{enumerate}
All of possibilities listed above appear in the literature
\cite{DN,Petro,HHRS,Bars}.
However, the possibility 5) does not work: even if we include
non-unitary representations, our argument in
Sec. 4 does not change very much and we can easily find physical
states with negative norms.
We will discuss the possibility 1) in the next subsection,
which is different from previously discussed ones. But in this section, we
first make comments on
the other consistency conditions after making some
general remark.

The basic physical consistency conditions for a string theory are not many.
In general, as a sensible physical theory, we must require Lorentz
invariance, a positive inner product for the observable Hilbert space and
the unitary transition amplitude. There are few in number, but these in turn
imply
various consistency conditions such as world-sheet diffeomorphism and Weyl
invariance, the absence of negative norm states, unitarity (closure of
OPE) and modular invariance.  Even though the absence of a tachyon might
also be added to the list, the presence of a tachyon in the bosonic string
does not indicate any fundamental inconsistency
in the theory.\footnote{See however Ref.~\cite{Joe},
which might imply that the bosonic
string does not exist nonperturbatively.}
Also, for modular invariance, it is sufficient to check associativity of
OPE and modular invariance of the one-point amplitude at one-loop
\cite{Sonoda}.

It does not sound an easy job for a string theory to satisfy all these
requirements. However, there is a common belief that a world-sheet anomaly
(either local or global) always leads to a spacetime anomaly.\footnote{
Some works on this theme are as follows: the connection of the
modular invariance and spacetime anomalies are discussed in \cite{CaiJoe}
(for the type I) and \cite{SWarner} (for the type II and the heterotic
string); the connection of the modular invariance and unitarity are
discussed in \cite{CaiJoe,Big}.}
So, a string theory is likely to be automatically consistent once world-sheet
anomalies are removed. If this is true even for curved backgrounds, the
most plausible solution to our ghost problem is the possibility 2).
This might be related to 1).
However, the modular invariance for a string theory in
a curved spacetime is a hard problem and not well understood.

\parbigskipn
{\it Closure of OPE}
\parsmallskipn
Unitarity requires closure of OPE,
and fusion rules are determined by tensor products of the underlying
primaries and by null states in Kac-Moody and Virasoro module.
Here we consider constraints on the fusion rule from tensor
products of the $ \cSLtwo $ representations.

Tensor products and the Clebsch-Gordan coefficients of
the $ \SLtwo $ representations including non-unitary ones
are discussed in \cite{CL}. Since we are dealing only
with the unitary representations, the problem is simple
and we can use the results in the literature.
We have summarized tensor products of the unitary representations
in Appendix A. We find that the tensor products
are closed
if the content of the operators is given by
(i) only the highest (or the lowest) discrete series,
or (ii)  the highest, lowest discrete series and
the principal continuous series, so that the addition and subtraction
of the $j$-values are closed mod
$ \bfZ $. Once we add the complementary series, we have to
include all the other unitary series.
These are the necessary conditions for the closure of
the OPE.

\parbigskipn
{\it Partition function and modular invariance}
\parsmallskipn
{}From the spectrum in Sec. 3, we get
\eqabegin
 L_0 - \Lbar_0  &=& - nm + N - \Nbar \comma \nn \\
 L_0 + \Lbar_0  &=&  \frac{-2 j(j+1)}{k-2} + N + \Nbar
      - n \lb \frac{k}{2} \deltap ^2 n - 2 \deltap J_R + m  \rb
     \period
\eqaend
The partition function diverges
since the Casimir $ -j(j+1)$, $ J_R $ and two integers $ n, m $ can
take arbitrarily large or small values.
In Minkowski spacetime, we can avoid the divergence
of the partition function by the Wick rotation, but
we have no analogue in our case. Furthermore, our Kac-Moody
module is restricted to the states of the form (\ref{KMmodule}), so
we have to take this into account in the character calculation.

One resolution to this problem might be to find a subclass of the spectrum
and/or to develop an analogue
of the Wick rotation so that we get a finite and
modular invariant partition function. This
might also solve the ghost problem.
For compact group manifolds \cite{GW}, the spectrum is
restricted to integrable representations
of the Kac-Moody algebra, so that we can get modular invariant partition
functions. Fields in non-integrable representations
decouple in correlators. However, the argument depend largely upon
compactness, so we have to take different strategies for non-compact
cases. So far, there is no general argument, but for the $ \SLtwo $ theory,
there are a few attempts\cite{HHRS,Huitu}. Besides group manifolds,
partition functions of string theories on curved spacetime are
discussed in \cite{RT}.

\subsection{Discrete symmetries}
One possibility to consistently truncate the spectrum
is further orbifolding besides that with respect to
$ \varphi \sim \varphi + 2 \pi $. As we will see,
only  part of the $ \cSLtwo $ manifold is necessary to describe
three dimensional black holes. Since we have started with
the $ \cSLtwo $ WZW model, the redundant part of the manifold
should be divided away by orbifolding. In this subsection, we will discuss
the relevant discrete symmetries.

In Appendix A, we see that the $ \SLtwo $ manifold contains
sixteen domains denoted by $ \pm D^\pm_i $ $ \ (i = 1$-4$)$.
One correspondence between Region I-III and these domains is
\eqabegin
   && \mbox{Region I  } =  D_1^+ \comma \quad
      \mbox{Region II  } = D_2^- \cap \lb -D_3^+ \rb \comma \quad
      \mbox{Region III  } = -D_4^- \period
\eqaend
Here we have taken a parametrization in Region II and III slightly different
from the one in Sec. 2, but the geometry is
the same. Thus we need only the universal covering
of the region
$ \Omega_1 \equiv D_1^+ \cap D_2^- \cap \lb -D_3^+ \rb \cap \lb -D_4^- \rb $
to get the black hole geometry,
as long as we do not consider its maximal extension.
Now let us define two transformations
by
\eqabegin
   \begin{array}{llllll}
  T_1 : &  g \to & g' = - g \comma
   &  &
    &  \\
  T_2 : &  g \to & g' = \calB g
   &  \mbox{ in } \pm D^\pm_{1,2}
  \comma &
   g' =  - \calB g
    &  \mbox{ in } \pm D^\pm_{3,4}
   \comma
    \end{array}
\eqaend
where $ \calB $ is given by (\ref{Bauto})
and called Bargmann's automorphism of $ \SLtwo $.
$ T_{1,2} $ have the properties
\eqabegin
  \begin{array}{ll}
   T_1^2 = T_2^2 = 1 \comma &  \\
    T_1 : \ \Omega_{1 (2) } \to - \Omega_{1(2) } \comma \quad &
    T_2 : \ \Omega_{1 (2) } \to  \Omega_{2 (1) } \comma
  \end{array}
\eqaend
where $ \Omega_2  =  \lb  D_1^- \cap D_2^+ \cap  D_3^- \cap  D_4^+ \rb $.
Note that $ \pm \Omega_{1,2} $ cover all sixteen domains of $ \SLtwo $
and have no overlap among them. Moreover we can obtain
the black hole geometry from each of the four sets as in Sec. 2.
Thus we can divide $ \SLtwo $ by the $ \bfZ _2 $ symmetries,
$ T_1 $ and $ T_2 $, in order to drop redundant regions.

There is one more discrete symmetry.
This is related to
the problem of closed timelike curves.
Region I-III or
each of $ \pm \Omega_{1,2} $ includes
the region $ r^2 < 0 $ where closed timelike curves exist \cite{BTZ}.
This region corresponds to part of $ -D_4^- $ in $ \Omega_1 $
for the rotating case or the whole region for the non-rotating case.
Although we have no symmetry to remove this region only,
it is possible to drop it together with
the region $  (\rp^2 + \rmi^2 )/2 > r^2 > 0 $. The region
$  (\rp^2 + \rmi^2 )/2 > r^2  $ corresponds to
$ \lb -D_3^+ \rb \cap \lb -D_4^- \rb $ in $ \Omega_1 $, so we have only
to find a symmetry between $ D_1^+ \cap D_2^+ $
and $ \lb -D_3^+ \rb \cap \lb -D_4^- \rb $.
The symmetry is easy to find
in coordinates $ ( \hatt , \hatphi , \hatr ) $.
Let us define a $ \bfZ _2 $ transformation by
\eqabegin
  T_3 : && \lb \hatt \comma \hatphi \comma \hatr^2 - 1/2 \rb \quad
    \to  \lb \hatphi \comma \hatt \comma - (\hatr^2 - 1/2 ) \rb
  \period
\eqaend
Then
by recalling that the geometry is given by (\ref{GBhat}),
we find that the metric and
the antisymmetric tensor are invariant under $ T_3 $.
This symmetry maps any point in $  D_1^+ \cap D_2^+ $ $(\hatr ^2 > 1/2 )$
to one in $ \lb -D_3^+ \rb \cap \lb -D_4^- \rb $ $(\hatr ^2 < 1/2 )$ and
vice versa. Thus we can truncate both the spectrum and the region
with closed timelike curves by the orbifolding with respect to
$ T_3 $ at the expense of the additional dropped region.
Notice that part of $ T_3 $, $ \hatr ^2 \to 1 -  \hatr ^2 $
or $ \hatt \leftrightarrow \hatphi $,  has
already appeared in the discussion of the T-duality in Sec. 5.

\parbigskipn
{\Large\bf Acknowledgements }
\parmedskipn
We would like to thank S. Hirano, G. Horowitz, H. Ishikawa, M. Kato,
Y. Kazama, Y. Matsuo and N. Sakai
for useful discussions.
We would especially like to thank T. Oshima for useful discussions on
the representation theory of $ \SLtwo $ and J. Polchinski for useful
discussions and
comments on the draft of this paper.
This work was supported in part by
JSPS Research Fellowship for Young Scientists No. 07-4678
and No. 06-4391.
%
%%%%%%%%%%%%%%%%%%%%
\appendix
%%%%%%%%%%%%%%%%%%%%
%
%%%%%%%%%%%%%%%%%%%%
%    appendix A
%%%%%%%%%%%%%%%%%%%%
%
\section{Representations of $ \SLtwo $}
\appendixnumbering{A}
In this appendix, we briefly summarize the
representation theory of $ \SLtwo $
(and its universal covering group $ \cSLtwo $ ) and collect its useful
 properties for discussions in this paper. For a review,
see \cite{VK} and \cite{Bargmann}-\cite{Wybourne}.
\subsection{$\SLtwo$}
\subsubsection{Preliminary}
The group $\SLtwo$ is represented by real matrices
\eqabegin
  && g = \matrixii{a}{b}{c}{d} \comma \qquad ad -  bc = 1.
\eqaend
It has one-parameter subgroups
\eqabegin
   \Omega_a &=& \lmb  g_a (t) = \ e^{ -i t \tau^a} \rmb \comma
  \qquad a = 0, 1, 2\comma
\eqaend
where
\eqabegin
  \begin{array}{llllll}
  \tau^0 &  & =
 {\displaystyle -\frac{1}{2} \sigma_2 }
    & \to  &
  g_0 (t) & =  \matrixii{ \cos t/2 }{ \sin t/2 }{ - \sin t/2}{ \cos t/2}
    \comma \\
  \tau^1 & & =
 {\displaystyle \frac{i}{2} \sigma_1 }
    & \to  &
   g_1 (t) & = \matrixii{\cosh t/2}{\sinh t/2 }{\sinh t/2}{\cosh t/2}
    \comma \\
 \tau^2 &  & =
{\displaystyle \frac{i}{2} \sigma_3 }
    & \to  &
    g_2 (t) & =  \matrixii{ e^{t/2}}{0}{0}{ e^{-t/2}}
    \comma
  \end{array}
\eqaend
where $ \sigma_i \ (i = 1$-$3)$ are the Pauli matrices.
In $ \Omega_0 $, $ g_0 (0) $ and $ g_0(4 \pi) $ represent the same point
and $ g_0(t), \  t \in [0,4 \pi) $
traces an uncontractable loop in $ \SLtwo $.
If we unwrap this loop and do not identify  $ g_0 (0) $ and $ g_0 (4 \pi ) $,
we get the universal covering group $ \cSLtwo $.
$ \tau^a (a = 0,1,2) $ have the properties
\eqabegin
  \left[ \tau^a , \tau^b \right] &=& i \epsilon^{ab}_{\ \ c} \tau^c
 \comma  \qquad
 \Tr \left( \tau^a \tau^b \right) = - \half \eta^{ab} \comma
    \label{sltwo}
\eqaend
where $ \eta^{ab} = \mbox{ diag }(- 1, 1, 1) $.
$ \tau^a $ form a basis of $ \sltwo $.

$ \SLtwo $ is isomorphic to $ SU(1,1) $
(and so is $ \sltwo $ to $ su(1,1) $).
The isomorphism is given by
\eqabegin
  && \tilg = T^{-1} \ g \ T \comma \qquad
T = \frac{1}{\sqrt{2}} \matrixii{1}{i}{i}{1}
   \label{iso} \comma
\eqaend
where $ \tilg \in SU(1,1) $ and $ g \in \SLtwo $.
Note $ \tilg _0 $ is diagonal in $ SU(1,1) $, while so is $ g_2 $
in $ \SLtwo $.

\subsubsection{Parametrization}
Any matrix $ g $ of $ \SLtwo $, with all its
elements being non-zero, can be represented as
\eqabegin
   g &=& d_1 \ (-e)^{\epsilon_1}\  s^{\epsilon_2} \ p \ d_2.
\eqaend
Here, $\epsilon_{1,2} = 0 $ or $ 1 $; $ d_i = $ diag
$ ( e^{\psi_i/2}, e^{- \psi_i/2}) $ $ (i = 1,2 ) $;
\eqabegin
   -e &=&  \matrixii{-1}{0}{0}{-1} , \qquad
    s = \matrixii{0}{1}{-1}{0} ,
\eqaend
and $p$ is one of the following matrices:
\eqabegin
  p &=& g_1(\theta)
  \comma \qquad \ -\infty < \theta < + \infty \comma \nn \\
  p &=&  g_0(\theta)
  \comma \qquad -\pi /2 < \theta < + \pi /2
   \period
\eqaend
Thus, $ \SLtwo $ has eight domains given by
\eqabegin
 D_1 &=& \lmb A_1
  = \matrixii{ e^{\phi} \cosh \theta/2}{ e^{\psi} \sinh \theta/2}
{ e^{-\psi} \sinh \theta/2}{ e^{-\phi} \cosh \theta/2}
  \comma \  -\infty <  \theta < + \infty \rmb  \comma \nn \\
 D_2 &=& \lmb  A_2
  = \matrixii{ e^{\phi} \cos \theta/2}{ e^{\psi} \sin \theta/2}
{ - e^{-\psi} \sin \theta/2}{ e^{-\phi} \cos \theta/2}
   \comma
   \ - \frac{\pi}{2} < \theta < + \frac{\pi}{2}  \rmb  \comma \nn \\
D_3 &=& \lmb A_3
  = \matrixii{ - e^{\phi} \sin \theta/2}{ e^{\psi} \cos \theta/2}
{ - e^{-\psi} \cos \theta/2}{ - e^{-\phi} \sin \theta/2}
   \comma
   \ - \frac{\pi}{2} < \theta < + \frac{\pi}{2}  \rmb  \comma \\
 D_4 &=& \lmb   A_4
  = \matrixii{ e^{\phi} \sinh \theta/2}{ e^{\psi} \cosh \theta/2}
{ - e^{-\psi} \cosh \theta/2}{ - e^{-\phi} \sinh \theta/2}
  \comma \ -\infty <  \theta < + \infty \rmb  \comma \nn \\
 - D_i &=& \lmb  - A_i \rmb \quad ( i = 1 \sim 4 ) \nn
\comma
\eqaend
where $  \ -\infty < \phi \comma \psi <   + \infty$.
We can further divide these domains according to the sign
of  $ \theta $. We denote the domains with positive $ \theta $ by
$ \pm  D_i^+ $ and those with negative $ \theta $ by $ \pm D_i^- $.

When a matrix element of $ g $ is zero, it is for example written by
$ \matrixii{a}{0}{b}{a^{-1}} $.
Taking appropriate limits of $ \pm A_i $ yields such a matrix.

\subsection{Unitary representations}
Let us denote the generators of
$ \sltwo $ by $ J^a $ and
consider the basis given by $ I^0 = J^0 $ and $ I^\pm = J^1 \pm i J^2 $.
In this basis, the nontrivial commutation relations read
\eqabegin
  \lbb I^0 , I^\pm \rbb &=& \pm I^\pm \comma \quad
 \lbb I^+ , I^- \rbb = - 2 I^0 \period  \label{ellipticcom}
\eqaend
This basis is natural from the $ su(1,1) $ point of view because
$ I^0 $ corresponds to diagonal elements and $ I^\pm $ are regarded
as ladder operators as in $su(2)$. Using this basis,
we can classify all unitary representations of $ \sltwo $ and hence
those of $ \SLtwo $ and $ \cSLtwo $ \cite{Bargmann},\cite{VK,DLP}.
There are five
classes of the unitary representations of $ \sltwo $ which are labeled by
the Casimir $ \bfC = \eta_{ab} J^a J^b $, $ I^0 $ and
a parameter $ m_0 \in [ \ 0 , 1 ) $:
\begin{enumerate}
 \item Principal continuous series $ T^P_\chi $ : \quad
    Representations realized in
     $ \left\{  \ket{j,m} \right\} $, $ m = m_0 + k $,
     $ 0 \leq m_0 < 1$, $k \in {\bf Z}$ and $ j = -1/2 + i \nu $,
     $ 0 < \nu $.
 \item Complementary (Supplementary) series $ T^C_\chi $ : $ \ $
   Representations  realized in
   $ \left\{  \ket{j,m} \right\}$, $ m = m_0 + k $,
     $ 0 \leq m_0 < 1$, $k \in {\bf Z}$,
     and $ \min \left\{ - m_0, m_0 -1 \right\} < j \leq -1/2 $.
 \item Highest weight discrete series  $ T^H_j $ : \quad
   Representations  realized in
     $\left\{  \ket{j,m} \right\} $, $ m = M_{max} - k$,
     $k \in {\bf Z}_{\geq 0} $ and
     $ j = M_{max} \leq -1/2  $ such that $ I^+ \ket{j,j} = 0 $.
 \item Lowest weight discrete series $ T^L_j $ : \quad
   Representations realized in
     $\left\{  \ket{j,m} \right\} $, $ m = M_{min} +  k$,
     $k \in {\bf Z}_{\geq 0} $ and
     $ j = - M_{min} \leq -1/2 $ such that $ I^- \ket{j,-j} = 0 $.
 \item Identity representation : \quad  The trivial representation
      $ \ket{-1,0} $.
\end{enumerate}
Here, $ \chi $ is the pair $(j,m_0)$;
$ {\bf Z}_{\geq 0} $ refers to  non-negative integers; and
we have denoted the value of
$ \bfC $ by $ -j(j+1) $. Note that $ j $ need not be real although
$ -j(j+1) $ should be and that we can restrict $ j $ to Im $ j > 0 $ for 1)
and $ j \leq -1/2 $ for  the others because $ j $ and $ -(j+1) $
represent the same Casimir.

Unitary representations of $ \cSLtwo $ are realized in the same space
$ \{ \ket{j,m} \}$.
For $ \SLtwo $,  the parameters are
further restricted to $ m_0 = 0 , 1/2 $ in  1), $ m_0 = 0 $ in  2)
and $ j = $ (half integers) in 3) and 4). We will use the same
notations as in $ \sltwo $.

{}From the harmonic analysis on $ \cSLtwo $,
a complete basis
for the square integrable functions on $ \cSLtwo $ is given by
the matrix elements of the principal continuous series, the highest
and lowest weight discrete series.

\subsection{Tensor product}
Because we have various unitary representations, the decomposition of tensor
products is more complicated than $ SU(2) $. Basic strategy
to get the decomposition is to decompose the tensored representation spaces
into the eigenspaces of the Casimir \cite{Pukanszky,Neunhoffer}.
We are interested in tensor products among $ T^P_\chi $ and $ T^{H,L}_j $.
For $ \SLtwo $, the decompositions are given as follows
\cite{VK,MR} :
\parmedskipn
1) For two discrete series of the same type,
\eqabegin
  T^{L,H}_{j_1} \ \otimes \ T^{L,H}_{j_2} &=&
   \sum_{n=0}^\infty \ \oplus \ T^{L,H}_{j_1 + j_2 - n }
 \period
\eqaend
\noindent
2) For two discrete series of different types,
\eqabegin
  T^L_{j_1} \ \otimes \ T^H_{j_2} &=&
  \int_0^\infty \ T^P_{(-1/2 + i \rho, m_0)} \ d \mu (\rho) \ \oplus \
   \sum_{ j = -m_0 -1}^{j_1-j_2} \ \lb  T^L_{j}  \oplus  T^H_{j} \rb
 \comma
\eqaend
where $ m_0 = j_1 - j_2 $ mod $ \bfZ $
and $ d \mu (\rho) $ is a continuous measure.
We have assumed
$ j_2 \geq j_1 $, but
the opposite case is obtained similarly.
We remark that $ j \leq -m_0 - 1 $ and the identity representation
does not appear in the right-hand side \cite{MR}.\footnote{
In \cite{HB}, it is claimed that the identity
representation does appear as an exceptional case. In our understanding,
they show just the existence of the solution to
the recursion equation for the Clebsch-Gordan coefficients.}

\parmedskipn
3) For a discrete and a principal continuous series,
\eqabegin
T^{L,H}_{j_1} \ \otimes \ T^P_{(-1/2 + i \rho', m'_0)} &=&
  \int_0^\infty \ T^P_{(-1/2 + i \rho, m_0)} \ d \mu (\rho) \ \oplus \
   \sum_{ j = -m_0 -1}^{-\infty} \  T^{L,H}_{j}
 \comma
\eqaend
where $ m_0 = m'_0 + j_1 $ mod $ \bfZ $.

\parmedskipn
4) For two continuous series,
\eqabegin
  && T^P_{(-1/2 + i \rho', m'_0)} \ \otimes
    \ T^P_{(-1/2 + i \rho'', m''_0)}  \label{CC} \\
  && \quad =
 \int_0^\infty \ T^P_{(-1/2 + i \rho, m_0)} \ d \mu_1 (\rho) \ \oplus \
  \int_0^\infty \ T^P_{(-1/2 + i \rho, m_0)} \ d \mu_2 (\rho) \ \oplus \
   \sum_{ j = -m_0 -1}^{-\infty}
   \ \lb  T^L_{j} \oplus  T^H_{j} \rb
  \comma \nn
\eqaend
where $ m_0 = m'_0 + m''_0 $ mod $ \bfZ $.

\parmedskipn
The decomposition is determined essentially by local properties of the group
as is clear from the consideration of tensor products of
$ \sltwo $. Thus the decompositions for $ \cSLtwo $ are obtained
by continuing the value of $ m_0 $ and $ j $.

For completeness, we mention tensor products including the complementary
series \cite{Pukanszky,Neunhoffer}.
The tensor product of a principal and a complementary
series, or that of two complementary series is decomposed into principal
and discrete series like (\ref{CC}). In the latter case,
one complementary series may appear additionally. The tensor product of
a complementary and a discrete series is similar to that of a principal
and a discrete series \cite{Neunhoffer}.

The Clebsch-Gordan coefficients have been discussed in
\cite{HB,MR}, \cite{VK,Wybourne}, \cite{CL}.

\subsection{Representations in the hyperbolic basis}
In Appendix A.2, we have discussed the representations in the basis
diagonalizing
$ J^0 = I^0 $ which is the compact direction of $ \SLtwo $.
We can also consider bases diagonalizing
$ J^2 $ or $ J^- = J^0 - J^1 $ which are non-compact
directions \cite{VK}, \cite{MR}, \cite{KMS}-\cite{BP}, \cite{DVV}.
The generators $ J^0 $, $ J^2 $ and $ J^- $
are called elliptic, hyperbolic and parabolic respectively. One
outstanding feature of non-compact generators is that they have
continuous spectra. In the rest of this appendix, we will
concentrate on representations in the
hyperbolic basis.

In terms of $ J^\pm \equiv J^0 \pm J^1 $ and $ J^2 $, the commutation
relations (\ref{sltwo}) are given by
\eqabegin
  \lbb J^+ , J^- \rbb &=& -2 i J^2 \comma \quad
  \lbb J^2 , J^\pm \rbb \ = \  \pm i J^\pm
   \label{comrel} \period
\eqaend
The latter equation indicates that
the ladder operators $ J^\pm $ change the eigenvalue of $ J^2 $ by
$ \pm i$. This seems to contradict the Hermiticity of $ J^2 $.
However, this is not the case \cite{KMS}.

In general, the eigenvalue of an Hermite operator with {\it continuous }
spectrum need not be real \cite{AFIO}, but for our purpose it is
convenient to choose spectrum with real values.
Thus, we use the basis given by
$ \{ \ket{ \lambda } \} $, where $ \lambda $ is the eigenvalue of
$ J^2 $ and runs through all the real number.
For the principal continuous and the complementary series,
the eigenvalue of $ J^2 $ has multiplicity two. Thus the basis has an
index $ \pm $ to distinguish them and is given
by $ \{ \ket{ \lambda } _\pm \} $.
In the remainder of this section, we omit
this and the other indices to specify representations such as $ j , m_0 , L $
and $ H $.
In the above basis,
an element (a state) of the representation space is given by a
``wave packet"
\eqabegin
  \ket{ \phi } &=&
    \int_{-\infty}^\infty d \lambda \ \phi (\lambda) \ket{\lambda} \comma
    \quad
   \Vert \ \phi \ \Vert ^2 \ = \ \int_{-\infty}^\infty d \lambda  \
    \abs{\phi (\lambda)} ^2 \ < \ \infty .
\eqaend
This is analogous to a state in  field theory
where one uses a plane wave basis in infinite space.
Then the generators act on the state as
\eqabegin
  J^2 \ket{\phi} &=&
     \int_{-\infty}^\infty d \lambda \ \lambda \phi (\lambda) \ket{\lambda}
 \comma \nn \\
   J^+ \ket{\phi} &=&  \int_{-\infty}^\infty d \lambda \
    f_+ ( \lambda ) \phi (\lambda - i) \ket{\lambda} \comma  \label{J+-}\\
  J^- \ket{\phi} &=&  \int_{-\infty}^\infty d \lambda \
    f_- ( \lambda + i ) \phi ( \lambda + i ) \ket{ \lambda }
   \period  \nn
\eqaend
$ f_\pm $ play the role of the matrix elements in this basis.
{}From the
above action,
the commutation rules are realized if
\eqabegin
    && f_+ (\lambda )  f_- (\lambda )
     -  f_- (\lambda + i )  f_+ (\lambda + i ) = - 2 i \lambda \period
  \label{fpfm}
\eqaend
An eigenstate $ \ket{\lambda '} $ is obtained in the limit
$ \phi (\lambda ) \to \delta ( \lambda - \lambda ') $.

It is possible to introduce $ \ket{\lambda \pm i} $ and write the action
of the generators as
\eqabegin
   J^+ \ket{\phi} &=&  \int_{-\infty}^\infty d \lambda \
    f_+ ( \lambda + i) \phi (\lambda) \ket{\lambda + i} \comma \nn \\
 J^- \ket{\phi} &=&  \int_{-\infty}^\infty d \lambda \
    f_- ( \lambda ) \phi (\lambda) \ket{\lambda - i} \comma
  \label{lambdapmi} \\
   J^+ \ket{ \lambda } &=& f_+ ( \lambda + i) \ket{\lambda + i}
  \comma \quad
  J^- \ket{ \lambda } \ = \ f_-  ( \lambda ) \ket{\lambda - i}
 \period \nn
\eqaend
In this way, we can formally consider eigenstates $ \ket{ \lambda \pm i } $.
However, we should always understand them
in the sense of (\ref{J+-}). Note that $ \ket{\lambda \pm i} $ can be
``expanded" by the original basis $\{ \ket{ \lambda } \}$, where
$ \lambda \in \bfR $.

Now let us consider the matrix elements of $ J^\pm $. In the elliptic basis,
we get the matrix elements of  $ I^\pm $ by
evaluating the commutation relation $ [ I^+ , I^- ] = - 2 I^0 $ between
eigenstates of $ I^0 $. In the hyperbolic basis, this method
dose not work because $ ( J^\pm ) ^{\dag} = J^\pm $.
First, note that the Casimir takes the form
\eqabegin
  \bfC &=& \eta_{ab} J^a J^b \nn \\
         & = &  J^2 (J^2 + i) - J^- J^+ = J^2 (J^2 - i) - J^+ J^-
 \label{casimirh} \period
\eqaend
The condition (\ref{fpfm}) has the solution
 $ f_+ (\lambda )  f_- (\lambda )  = \lambda ( \lambda - i ) - c$,
where $ c $ is a constant. Moreover,
evaluating the Casimir on an eigenstate of $ J^2 $ leads to
$ c = - j(j+1) $, i.e.,
\eqabegin
  f_+ (\lambda )  f_- (\lambda )  &= & \lambda ( \lambda - i ) + j(j+1)
 \  \equiv  \  d^2 (j, \lambda - i)
  \period
\eqaend
We cannot determine $ f_+ $ or $ f_- $ separately
without additional conditions.
Consequently, we find that
\eqabegin
  J^+ J^-  \ket{j;\lambda} &=& d^2 (j, \lambda - i) \ket{j;\lambda}
    \comma  \quad
  J^- J^+  \ket{j;\lambda}  \  = \ d^2 (j, \lambda ) \ket{j;\lambda}
 \period \label{Jpm}
\eqaend
Note that $ d^2 (j, \lambda - i) = \overline{ d^2 (j, \lambda ) }$.

In the elliptic basis, the commutation
relations are given by (\ref{ellipticcom}) and the Casimir is by
\eqabegin
  && \bfC = - I^0 ( I^0 + 1) + I^- I^+ = - I^0 ( I^0 - 1) + I^+ I^-
 \period \label{ellipticC}
\eqaend
{}From them, we find that
the actions of $ I^\pm $ on an eigenstate of $ \bfC $ and $ I^0 $
are
\eqabegin
   && I^- I^+ \ket{j;m} = \tilde{d}^2 (j,m) \ket{j;m} \comma \quad
      I^+ I^- \ket{j;m} = \tilde{d}^2 (j,m-1) \ket{j;m} \comma
  \label{Ipm}
\eqaend
where
$ \tilde{d}^2 (j,m) = -j(j+1) + m (m +1)$.
So, we see that (\ref{ellipticcom}), (\ref{ellipticC}) and (\ref{Ipm})
are related to
the corresponding equations in the hyperbolic basis
by ``analytic continuation"  $ I^\pm \to i J^\pm $ and
$ I^0 \to -i J^2 $ \cite{KMS}.

\subsection{Matrix elements}
By explicit realization of the representations in spaces of function, we
can calculate the matrix elements of $ \SLtwo $.
Here we consider the matrix elements in the hyperbolic basis
\cite{VK}, \cite{Mukunda,BP}, \cite{DVV}.

First, let us discuss the principal continuous series
$ T^P_\chi $ of $ \SLtwo $.
This representation is realized in a space of functions
on a real axis, $ \Ichi $.
The action of
the group element $ \matrixii{a}{b}{c}{d} \in \SLtwo $ and the inner
product are given by
\eqabegin
  \left(  T^P_\chi (g) f \right) (x) &=& \abs{b x + d} ^{2j}
     \mbox{ sign }^{2m_0} (bx + d) \ f \left(\frac{ax + c}{bx + d} \right)
    \label{repIchi} \comma \\
 \lb f_1(x) , f_2(x) \rb &=&
   \int_{-\infty}^\infty d x \ \overline{f_1 (x)} f_2 (x)
  \label{inprod} \period
\eqaend
Then we find that
\eqabegin
 \psi^\chi_{\lambda \pm} (x) & \equiv &
      \frac{1}{\sqrt{2 \pi}} x^{- i \lambda + j } \theta (\pm x )
      \comma \quad \lambda \ \in {\bf R} \label{psi2} \comma
\eqaend
form an orthonormal basis diagonalizing the action of $ J^2 $, namely
\eqabegin
  \lb \psi^\chi_{\lambda \epsilon} (x) , \psi^\chi_{\mu \epsilon'} (x) \rb
   &=& \delta_{\epsilon \epsilon' } \delta (\lambda - \mu )  \comma \\
\lbb T^P_\chi \lb g_2 (t) \rb \psi^\chi_{\lambda \pm} \rbb (x) &=&
   e^{-it \lambda } \psi^\chi_{\lambda \pm} (x) \comma
   \quad g_2 (t)\ \in \Omega_2 \comma
\eqaend
where $\epsilon \comma \epsilon' = \pm $.
$ \psi^\chi_{\lambda \pm} $ correspond to $ \ket{ \lambda } _\pm $ in the
previous subsection and are not elements in $ \Ichi $.

We can calculate the
matrix elements in the basis (\ref{psi2}) using (\ref{repIchi}) and
(\ref{inprod}). For example, for $ t > 0 $
we have
\eqabegin
    \D{P} ^{\chi}_{\lambda +,\lambda' +} \lb g_1(t) \rb
  & = & \frac{1}{2 \pi } B \lb \mu, -\mu' -2j \rb
 \frac{\cosh^{2j + \mu + \mu'} t/2 }{ \sinh^{\mu + \mu' } t/2 }
    \label{g1++} \\
   &&  \qquad \qquad \times F \lb \mu, \mu'; - 2 j ; - \sinh^{-2} t /2 \rb
   \comma  \nn  \\
   \D{P} ^{\chi}_{\lambda -,\lambda' -} \lb g_1(t) \rb
 & = & \frac{1}{2 \pi } B \lb 1-\mu', \mu' -1 +2(j+1) \rb
 \frac{\cosh^{2j + \mu + \mu'} t/2 }{ \sinh^{4 j + 2 +  \mu + \mu' } t/2 }
    \nn \label{g1--} \\
   & &  \times
  F \lb \mu+ 2j + 1 , \mu' + 2j + 1 ; 2 j + 2 ; - \sinh^{-2} t /2 \rb
  \comma  \\
   \D{P} ^{\chi}_{\lambda \epsilon,\lambda' \epsilon'} \lb g_2(t) \rb & = &
   e^{-i t \lambda }\delta_{\epsilon \epsilon'}
   \delta ( \lambda - \lambda' )  \comma \label{g2}
\eqaend
where $ \mu^{(')} = i \lambda^{(')} - j $. $ F $ and $ B $ are
the hypergeometric function and the Euler beta function respectively.
For $ g_1(t)$,
$ \D{P} ^{\chi}_{\lambda -,\lambda' +} $ is given by a linear
combination of (\ref{g1++}) and (\ref{g1--}), and
$ \D{P} ^{\chi}_{\lambda +,\lambda' -} $ vanishes.

The matrix elements for the complementary series are obtained
by analytically continuing the value of $ j $ \cite{Mukunda}.

Let us turn to the discrete series $ T^L_j $. This is realized in a
space of analytic functions on
$ {\bf C}_+ $ (the upper half-plane).
(This can also be embedded in the principal continuous series.)
The action of $ g = \matrixii{a}{b}{c}{d} \in \SLtwo $ and the
inner product are given by
\footnote{$ j = -1/2 $ case needs special treatment, but the
matrix elements take the same forms as in  $ j < -1/2 $ cases
\cite{Bargmann,Mukunda}. }
\eqabegin
 \lb T_j^L (g) f \rb (w) &=& \lb b w + d \rb^{2j}
  f \lb \frac{a w + c}{b w + d} \rb \comma \\
 \lb f_1(w) , f_2 (w) \rb &=&
   \frac{i}{2 \Gamma (-2j-1)} \int_{\bfC _+} d w d \wbar \ y^{-2j-2}
   \overline{f_1 (w)} f_2 (w) \comma
\eqaend
where $ w = x + i y $ and $dw d \wbar= -2i dx dy $.
We then find that
\eqabegin
\varphi^j_{\lambda}(w) &=& \frac{1}{2^{(j+1)} \pi }
     e^{-\lambda \pi/2}
   \Gamma (-i\lambda -j) \  w^{-i \lambda + j}
  \comma \quad \lambda \ \in \bfR \comma
\eqaend
form an orthonormal basis diagonalizing $ J^2 $. Thus similarly
to the previous case (or using the fact that $ f(w) $ is determined by
its values on the semi-axis $ w = i y \ (y > 0)$), we can calculate the matrix
elements.
$  \D{L} ^{j}_{\lambda,\lambda'} \lb g_1(t) \rb $ is  the same up to
a numerical factor as (\ref{g1++}) and
$ \D{L} ^{j}_{\lambda,\lambda'} \lb g_2(t) \rb $ is given by
  (\ref{g2})  without $ \delta_{\epsilon \epsilon'}$.

For the highest weight series $ T^H_j $, we can get the matrix elements from
the lowest weight series. By utilizing an automorphism of $ \SLtwo $
called Bargmann's
automorphism of $ \SLtwo $
\eqabegin
 {\cal B} : && \matrixii{a}{b}{c}{d} \to \matrixii{a}{-b}{-c}{d}
 \comma \label{Bauto}
\eqaend
the matrix elements of the highest weight series
 are given
by \cite{VK,Mukunda}
\eqabegin
  \D{H} ^{j}_{\lambda,\lambda'} \lb g \rb &=&
 \D{L} ^{j}_{\lambda,\lambda'} \lb {\cal B} g \rb
 \period
\eqaend

All the matrix elements satisfy the
differential equation
\eqabegin
  \lbb  \Delta - j (j +1) \rbb  D^{j(\chi)}_{\lambda,\lambda'} \lb g \rb
   &=& 0
  \comma
\eqaend
where
$ \Delta $ is the Laplace operator on $ \SLtwo $ and they are characterized
essentially by local properties of $ \SLtwo $.
Hence, the matrix elements of $ \cSLtwo $ are obtained by continuing the values
of
$ j $ and $ m_0 $.
%
%%%%%%%%%%%%%%%%%%%%%%%
%   appendix B
%%%%%%%%%%%%%%%%%%%%
%
\section{Decomposition of the Kac-Moody module}
\appendixnumbering{B}
The Clebsch-Gordan decomposition
similar to $ su(2) $ holds for $ \sltwo $ $ (su(1,1)) $ in the
elliptic basis \cite{BOFW}. Their argument is valid in the hyperbolic basis as
well with
a slight modification.

Let $ V^a $ be a vector operator, i.e.,
\eqabegin
  \lbb J_0^a , V^b \rbb &=& i \epsilon^{ab}_{\ \ c} V^c \comma
\eqaend
and $ \ketp{j;\lambda} $ be an eigenstate of $ \bfC $ and $ J^2 $.
An example is $ V^a = J^a_{-1}$.
$  \ketp{j;\lambda} $ need not be a base state of the Kac-Moody module.
Let us consider states
\eqabegin
   V^+ J_0^- \ketp{j;\lambda} \comma \qquad
   V^- J_0^+ \ketp{j;\lambda} \comma \qquad
   V^2 \ketp{j;\lambda} \label{VJ}  \period
\eqaend
In the hyperbolic basis,
these states do not vanish in any unitary representation.
{}From (\ref{casimirh}),
the matrix elements of the Casimir with respect to
these states are
\eqabegin
  \bfC &= & \matrixiii{c + 2 i \lambda }{0}{i}{0}{c - 2 i \lambda}{-i}
{-2i d^2(j,\lambda -i )}{ 2 i d^2 (j,\lambda)}{ c-2 }
 \comma  \quad \mbox{where} \quad c = -j(j+1) \period
\eqaend
The trace and determinant in this subspace are given by
\eqabegin
  && \Tr \ \bfC = 3 c - 2 \comma \qquad \det \bfC = c^2 (c + 2)
 \period
\eqaend
It is easy to see that the state $ ( 1,1, -2 \lambda ) $ is an eigenvector
with the Casimir $ \bfC = -j(j+1) $. Then, the other eigenvalues are
$ -j(j-1) $ and $ -(j+1)(j+2) $.
Therefore, the states in (\ref{VJ}) are decomposed into the $ \sltwo $
representations with $j$-values $ j $ and $ j \pm 1 $.
The corresponding eigenvectors $ \psi_j $ and $ \psi_{j\pm1}$
are given by
\eqabegin
  \psi_j &=& ( 1,1, -2 \lambda )  \comma \nn \\
  \psi_{j-1} &=& \lb -(j + i \lambda ) \comma
    j - i \lambda \comma 2 i (j^2 + \lambda^2) \rb
   \comma \\
    \psi_{j+1} &=& \lb j + 1 - i \lambda \comma  -(j + 1 + i \lambda )
   \comma  2 i ( (j+1)^2 + \lambda^2) \rb   \period \nn
\eqaend
Note $ \psi_{j+1} $ is obtained
from $ \psi_{j-1} $
by the replacement $ j \to -j-1 $.

It is useful to remark
on the norm of states. Consider
representations where the Casimir operator is Hermitian. The representations
need not be unitary. Let $ \ket{ \Psi_1} $ and $ \ket{ \Psi_2 } $ be
eigenstates with the Casimir values $ c_1 $ and $ c_2 $ respectively.
Then by evaluating the matrix element
$ \lb \Psi_1 \comma \bfC \Psi_2 \rb
  = \lb \bfC \Psi_1 \comma  \Psi_2 \rb $, we get
\eqabegin
  && \lb  \bar{c}_1 - c_2 \rb \bra{\Psi_1} \semiket{\Psi_2} = 0
  \period
\eqaend
Therefore, for complex $ c_1 $ and $ c_2 $,
the norm vanishes when $ c_1 = c_2 $.
It can be non-zero only when $ c_1 $ and $ c_2 $ are complex conjugate.
Since extremal states constructed on a principal continuous series
have complex Casimir values (see Sec. 4),
they are physical states with zero norm.
On the other hand, $ \bra{E^+_N} \semiket{E^-_N} $
can be non-zero because their Casimir
values are complex conjugate.

%
%%%%%%%%%%%%%%%%%%%%%%
\baselineskip=2.5ex
%%%%%%%%%%%%%%%%%%%%%%
%
%%%%%%%%%%%%%%%%%%%%%%%
%   references
%%%%%%%%%%%%%%%%%%%%%%%
%

%
%%%%%%%%%%%%%%%%%%%%%%%%%%%%

\begin{thebibliography}{99}
%
%%%%% SL(2,R)/U(1) BH %%%%%
  \bibitem{Witten} E. Witten, Phys. Rev. {\bf D44} (1991) 314.
  \bibitem{MSW} G. Mandal, A. Senguputa and S. Wadia, Mod. Phys. Lett.
               {\bf A6} (1991) 1685.
  \bibitem{Giveon} A. Giveon, Mod. Phys. Lett. {\bf 31A} (1991) 2843.
  \bibitem{DVV} R. Dijkgraaf, E. Verlinde and H. Verlinde, Nucl. Phys.
               {\bf B371} (1992) 269.
  \bibitem{DN} J. Distler and P. Nelson, Nucl. Phys. {\bf B374} (1992) 123.
  \bibitem{CL} S. Chaudhuri and J. Lykken,
              Nucl. Phys. {\bf B396} (1993) 270.
  \bibitem{IKOS} K. Itoh, H. Kunitomo, N. Ohta and M. Sakaguchi,
               Phys. Rev. {\bf D48} (1993) 3793.
%%%%% unitarity proof %%%%%
  \bibitem{noghost}
    P. Goddard and C.B. Thorn, Phys. Lett. {\bf 40B}, 235 (1972);\\
    R. Brower, Phys. Rev. {\bf D6}, 1655, (1972);\\
    M. Kato and K. Ogawa, Nucl. Phys. {\bf B212}, 443 (1983);\\
    C.B. Thorn, Nucl. Phys. {\bf B286} (1987) 61;\\
    D. Ghoshal and S. Mukherji, Mod. Phys. Lett. {\bf A6} (1991) 939.
%%%%% SL(2,R) string %%%%%%
  \bibitem{BOFW} J. Balog, L. O'Raifeartaigh, P. Forg\'{a}cs and A. Wipf,
               Nucl. Phys. {\bf B325} (1989) 225.
  \bibitem{DLP} L. Dixon, J. Lykken and M. Peskin,
              Nucl. Phys. {\bf B325} (1989) 329.
   \bibitem{Petro} P. Petropoulos, Phys. Lett. {\bf B236} (1990) 151;\\
                  S. Hwang, Nucl. Phys. {\bf B354} (1991) 100.
  \bibitem{HHRS} M. Henningson, S. Hwang, P. Roberts and B. Sundborg,
               Phys. Lett. {\bf B267} (1991) 350.
  \bibitem{GH} P. Griffin and O. Hernandez,
                  Nucl. Phys. {\bf B313} (1991) 287.
  \bibitem{Huitu}  K. Huitu, Phys. Lett. {\bf B313} (1993) 75.
  \bibitem{BN} I. Bars and D. Nemeschansky,
                   Nucl. Phys. {\bf B348} (1991) 89.
  \bibitem{Bars} I. Bars, Phys. Rev. {\bf D53} (1996) 3308.
%%%%% string curved in curved b.g. %%%%%
  \bibitem{RT} J.G. Russo and A.A. Tseytlin,
               Nucl. Phys. {\bf B449} (1995) 91.
%%%%% BTZ BH %%%%%
  \bibitem{BTZ} M. Ba\~{n}ados, C. Teitelboim and J. Zanelli,
               Phys. Rev. Lett. {\bf 69} (1992) 1849;\\
              M. Ba\~{n}ados, M. Henneaux, C. Teitelboim and J. Zanelli,
              Phys. Rev. {\bf D48} (1993) 1506.
%%%%% string theory of BTZ BH %%%%
  \bibitem{HW} G.T. Horowitz and D.L. Welch, Phys. Rev. Lett.
              {\bf 71} (1993) 328.
  \bibitem{Kaloper}N. Kaloper, Phys. Rev. {\bf D48} (1993)  2598.
%%%%% review of the BTZ BH %%%
  \bibitem{Carlip} S. Carlip, Class. Quant. Grav. {\bf 12} (1995) 2853.
%%%%% exact metric %%%%%%%%%
  \bibitem{BST}  I. Bars and K. Sfetsos, Phys. Rev. {\bf D46} (1992) 4510;\\
                 A.A. Tseytlin, Nucl. Phys. {\bf B399} (1993) 601;
                 {\bf B411} (1994) 509.
%%%%% twisting  %%%%%%%
  \bibitem{GPS} S.B. Giddings, J. Polchinski and A. Strominger,
              Phys. Rev. {\bf D48} (1993) 5784.
%%%%% modular invariance in compact WZW %%%%
\bibitem{GW} D. Gepner and E. Witten, Nucl. Phys. {\bf B278} (1986) 493.
%
%%%%% Vilenkin %%%%%%
  \bibitem{VK} N.Ja. Vilenkin and A.U. Klimyk,
     {\it Representation of Lie Groups and Special Functions},
      (Kluwer Academic Publishers, Dordrecht, 1991).
%%%%% tachyon and scalar in BTZ BH %%%
  \bibitem{GL} K. Ghoroku and A.L. Larsen, Phys. Lett. {\bf B328} (1994) 28;\\
        M. Natsuume, N. Sakai and M. Sato, Mod. Phys. Lett.
              {\bf A11} (1996) 1467.
  \bibitem{IS} I. Ichinose and Y. Satoh, Nucl. Phys. {\bf B447} (1995) 340.
%%%%% T-duality %%%%%
  \bibitem{Buscher} T. Buscher, Phys. Lett. {\bf B194} (1987) 59;
                 {\bf B201} (1988) 466.
  \bibitem{GRV} M. Ro{\v c}ek and E. Verlinde,
              Nucl. Phys. {\bf B373} (1992) 630;\\
              A. Giveon and M. Ro{\v c}ek,
              Nucl. Phys. {\bf B380} (1992) 128.
%
%%%%% consistency for a string theroy %%%%
   \bibitem{Joe} J. Polchinski, Phys. Rev. Lett. {\bf 74} (1995) 638.
   \bibitem{Sonoda}
     H. Sonoda, Nucl. Phys. {\bf B311} (1988) 401; 417.
   \bibitem{CaiJoe}
          J. Polchinski and Y. Cai, Nucl. Phys. {\bf B296} (1988) 91.
   \bibitem{SWarner}
      A.N. Schellekens and N.P. Warner, Phys. Lett. {\bf 177B} (1986) 317;
      {\bf 178B} (1986) 339; Nucl. Phys. {\bf B287} (1987) 317.
   \bibitem{Big} J. Polchinski, Nucl. Phys. {\bf B307} (1988) 61;
     {\it Joe's Big Book of String} (Cambridge, to appear).
%%%%% representation theory of SL(2,R) %%%%%%
%%% general %%%%
  \bibitem{Bargmann} V. Barmann, Ann. Math. {\bf 48} (1947) 568.
  \bibitem{Sally} P.J. Sally, Jr., Mem. Am. Math. Soc. {\bf 69} (1967).
  \bibitem{Wybourne} B.G. Wybourne, {\it Classical groups for physicists}
      (Wiley, New York, 1974).
%%% tensor product %%%
  \bibitem{Pukanszky} L. Pukanszky,
         Trans. Am. Math. Soc. {\bf 100} (1961) 116.
  \bibitem{Neunhoffer} H. Neunh\"{o}ffer,
     Sitzungsber. Heidelb. Akad. Wiss. Math. Natur. Kl. {\bf 3} (1978) 173.
  \bibitem{MR} N. Mukunda and B. Radhakrishnan,
     J. Math. Phys. {\bf 15} (1974) 1320; 1332; 1643; 1656.
  \bibitem{HB} W.I. Holeman, III and L.C. Biedenharn, Jr.,
     Ann. Phys. (NY) {\bf 39} (1966) 1; {\bf 47} (1968) 205.
%%% in the hyperbolic bassis %%%
  \bibitem{KMS} J.G. Kuriyan, N. Mukunda and E.C.G. Sudarshan,
       J. Math. Phys. {\bf 9} (1968) 2100.
  \bibitem{LN} G. Lindblad and B. Nagel,
       Ann. Inst. Henri Poincar\'{e} {\bf XIII} (1970) 27.
%%% matrix elements in the hyperbolic basis %%%
  \bibitem{Mukunda} N. Mukunda, J. Math. Phys. {\bf 10} (1969) 2086; 2092.
  \bibitem{BP} A.O. Barut and E.C. Phillips,
       Commun. Math. Phys. {\bf 8} (1968) 52.
%%%%%% Hermitian operator with a continuum spectrum %%%
  \bibitem{AFIO} H. Arisue, T. Fujiwara, T. Inoue and K. Ogawa,
     J. Math. Phys. {\bf 22} (1981) 2055.
%
 \end{thebibliography}
\end{document}